\documentclass[preprint,12pt]{elsarticle}

\biboptions{sort&compress}
\usepackage{amssymb}
\usepackage{amsmath}
\usepackage{graphicx}
\usepackage{subcaption}
\usepackage{algorithm}
\usepackage{algorithmicx}
\usepackage{algpseudocode}
\usepackage{booktabs}
\usepackage{multirow}
\usepackage{textcomp}
\usepackage[section]{placeins}
\usepackage{xcolor}
\usepackage{etoolbox}

\newcommand{\rev}[1]{#1}

\AtBeginEnvironment{table}{\renewcommand{\baselinestretch}{1.1}\selectfont}
\newcommand{\panelcaption}[1]{\vspace{-15pt}\caption{#1}}
\ifdefined\pdfobjcompresslevel
\fi

\makeatletter
\let\ps@pprintTitle\ps@plain
\makeatother

\patchcmd{\pprintMaketitle}{\vskip18pt}{\vskip6pt}{}{}
\patchcmd{\pprintMaketitle}{\vskip36pt}{\vskip6pt}{}{}
\patchcmd{\pprintMaketitle}{\vskip10pt}{\vskip4pt}{}{}
\begin{document}

\begingroup
\renewcommand{\baselinestretch}{1}\selectfont
\begin{frontmatter}

\title{\rev{Reduced-order topology optimization of defect spectra in phononic crystals}}

\author[inst1]{Xinlin Xu\corref{cor1}}
\ead{xu.xinlin.k4@f.mail.nagoya-u.ac.jp}

\author[inst1]{Junji Kato}

\cortext[cor1]{Corresponding author.}

\affiliation[inst1]{organization={Department of Civil and Environmental Engineering, Graduate School of Engineering, Nagoya University},
            addressline={Furo-cho},
            city={Chikusa-ku, Nagoya},
            postcode={464-8601},
            state={Aichi},
            country={Japan}}

\begin{abstract}
\renewcommand{\baselinestretch}{1}\selectfont
\interlinepenalty=0
\widowpenalties 4 10000 10000 10000 0
\clubpenalty=10000
\rev{Defect modes localize vibrations within phononic-crystal bandgaps. Prescribing their frequencies alone does not ensure separation from competing modes. We use frequency-dependent selection weights to combine attraction of target-adjacent modes with repulsion of competing in-gap modes. Only the central defect cell is optimized within a previously designed, fixed host bandgap. Cell-level Craig--Bampton reduction retains physical interface coordinates, allowing the host contribution to be assembled once while the defect basis and matrices are updated. Across three numerical cases, the reduced-order method accelerates defect optimization by approximately 13--20 times. A comparison with attraction-only optimization shows that the combined objective moves competing modes away from the target and increases the minimum target-to-non-target frequency separation. Full-order reanalysis shows that the optimized modes remain close to the prescribed frequencies and localized around the defect.} A finite-array calculation shows enhanced response near the target relative to the defect-free host under the prescribed loading.
\par
\end{abstract}


\end{frontmatter}
\endgroup
\raggedbottom
\setlength{\emergencystretch}{2em}

\section{Introduction}
\label{sec:introduction}

Phononic crystals (PnCs) control elastic waves through periodic modulation of stiffness and mass density.
Their bandgaps suppress wave propagation over selected frequency intervals and thereby support vibration isolation and wave filtering \cite{kushwahaAcousticBandStructure1993,husseinDynamicsPhononicMaterials2014}.
However, many wave-control applications also require a controlled localized response at a selected frequency inside the attenuation range.
This requirement motivates the intentional introduction of defects into an otherwise periodic host structure.

When translational symmetry is broken by a point defect, discrete modes can emerge inside an otherwise forbidden host bandgap \cite{shenGeneralizedSupercellModel2024}.
The corresponding modal field is concentrated around the defect and decays evanescently into the surrounding periodic host \cite{joRevealingDefectmodeenabledEnergy2022}.
Such localized states have been exploited for narrowband piezoelectric energy harvesting \cite{joDesigningPhononicCrystalDefect2020},
while active defect implementations further enable electrical and magnetic tuning of narrowband defect responses \cite{joSteeringMultipleDefectBands2025,zhangMagneticallyTunableBand2025}.
Beyond conventional point defects, topological interface and edge states provide robust localized responses for sensing and reconfigurable switching \cite{maTopologicalInterfaceStates2025,bahramiTopologicalPhononicSwitch2025}.

Topology optimization generates material layouts for wave-control structures without prescribing inclusion or defect shapes in advance~\cite{dongInverseDesignPhononic2024}.
Sigmund and Jensen established topology optimization as a systematic tool for phononic bandgap design \cite{sigmundSystematicDesignPhononic2003}.
Subsequent host-bandgap design has expanded to target-centered and multi-material formulations \cite{wuTargetedfrequencyBandgapMaximization2025,nakayamaMultimaterialTopologyOptimization2025}.
Coupled vibroacoustic bandgaps have also been addressed through topology optimization \cite{coolMetamaterialDesignVibroacoustic2025}.
\rev{For three-dimensional acoustic metamaterials, Yan et al.~\cite{yanFloatingProjection2025} combined bandgap optimization with permeability constraints to maintain connected air and solid domains, and introduced defects to guide wave propagation.}
Data-driven methods further accelerate inverse design of host bandgaps \cite{zhaoDatadrivenTopologyOptimization2025,liuFastTopologyOptimization2024}.
\rev{Beyond bandgap width and center frequency, Chen et al.~\cite{chenRAG2026} used a random-forest model to design partial passbands and stopbands over prescribed wave-vector and frequency intervals. This formulation controls selected regions of the dispersion relation; the present defect problem instead concerns the placement and separation of in-gap modes at $\Gamma$.}

For defective PnCs, topology optimization has been used to create specified narrow passbands \cite{zhangNarrowbandFilterDesign2021}.
More recently, deep-learning inverse design has progressed from single-defect target-frequency filtering to double-defect configurations with two prescribed defect-band frequencies \cite{leeDeeplearningbasedFrameworkInverse2023,leeInverseDesignPhononic2025}.
\rev{Li et al.~\cite{liTunablePointDefect2025} optimized the surrounding cells and a movable point defect for post-fabrication resonance tuning. Here the host remains fixed during defect optimization, and the design objective controls the distribution of frequencies inside its bandgap.}

\rev{Frequency placement and separation from unwanted modes are related design requirements. In MEMS resonators, Giannini et al.~\cite{gianniniTargetEigenfrequencies2022} prescribed drive and sense frequencies together with separation from spurious modes, using model reduction to accelerate the design. For defective PnCs, the additional structure is a host bandgap that bounds the admissible defect spectrum.} Attracting target-adjacent modes alone leaves their separation from non-target modes uncontrolled, whereas indiscriminate repulsion may displace the desired resonance. \rev{We therefore treat all current $\Gamma$-point eigenvalues inside the host-bandgap interval collectively, assigning attraction and repulsion through frequency-dependent selection weights. The computational challenge is to evaluate this evolving spectrum repeatedly as the supercell size increases to reduce interactions between periodic defect images.}

\rev{Projection-based model reduction lowers the cost of repeated analyses of parameter-dependent systems~\cite{bennerSurveyProjectionBasedModel2015}. Bogomolny~\cite{bogomolny2010combined} used combined approximations for free-vibration optimization, and Delissen et al.~\cite{delissen2020resonant} studied modal truncation and augmentation for resonance-peak constraints. Li et al.~\cite{liReducedOrderComparative2021} compared Ritz-vector, Pad\'e, and second-order Krylov methods for broadband frequency-response optimization. Successive iteration of analysis and design also reduces the nested cost of phononic band-structure optimization~\cite{zhuTopologyOptimizationFramework2023}. Accuracy during design changes can be controlled through residual-based basis updates~\cite{xiaoOnTheFlyPCA2020}, trust-region safeguards~\cite{yanoGloballyConvergent2021}, or primal--dual reduction~\cite{qianDualReduction2022,xiaoPrimalDual2024}. Recent learning-based approaches predict initial topologies for physics-based refinement~\cite{giacominiAutoencoders2026} or reduced bases enriched through residual corrections~\cite{matrayGeometryAgnostic2026}. Here the exploitable structure is spatial: the design changes within one cell of an otherwise fixed supercell.}

\rev{Dynamic substructuring~\cite{kohSubstructuring2020}, Craig--Bampton (CB) reduction in vibroacoustic topology optimization~\cite{wangEfficientAcousticTopology2020}, and coupled finite-element/wave-based models~\cite{gooHybridAcoustic2017} establish ways to treat design and non-design regions. Park et al.~\cite{parkUnassembled2026} further developed partitioned stiffness and flexibility formulations for unassembled models. Component-wise reduction reuses trained components in compliance and stress-constrained lattice optimization~\cite{mcbaneComponentWise2021,mcbaneStressConstrained2022}. For periodic wave analysis, Bloch mode synthesis~\cite{krattigerBlochModeSynthesis2014,krattigerGeneralizedBlochMode2018}, algebraic condensation~\cite{xiImprovingGeneralizedBloch2021}, and enhanced CB reduction~\cite{zhuImprovementGeneralizedBloch2023} reduce periodic wave models; enriched CB models and matrix interpolation also accommodate geometric parameter changes~\cite{mencikMatrixInterpolation2026}.}

\rev{The present formulation assigns attraction and repulsion to the current in-gap modes through frequency-dependent selection weights and exploits the localized design changes in the reduced assembly. The reduced assembly builds on standard component mode synthesis~\cite{craigCouplingSubstructuresDynamic1968,deklerkGeneralFrameworkDynamic2008} and component reuse in topology optimization~\cite{mcbaneComponentWise2021,mcbaneStressConstrained2022}, with the evolving defect cell supplying the iteration-dependent contribution. Physical interface coordinates allow the defect basis and matrices to be updated without rebuilding the host basis or its assembled contribution; the coupled reduced supercell eigenproblem is still solved at every iteration. An attraction-only comparison tests the spectral objective, fixed-topology full-order reanalysis assesses the reduction error, and matched full- and reduced-order optimizations measure the resulting design cost and target placement.}

\rev{Section~\ref{sec:formulation} defines the wave model. Section~\ref{sec:optimization} presents the spectral objective and reduced assembly, and Section~\ref{sec:sensitivity} describes mode recovery and sensitivity evaluation. Section~\ref{sec:numerical} reports the numerical results, followed by the conclusions. Standard filtering and material interpolation are collected in \ref{sec:implementation}.}

\section{\rev{Wave model and defect-spectrum definitions}}
\label{sec:formulation}

\rev{The finite-element and Bloch formulations define the eigenfrequencies used in the optimization and the host interval within which defect modes are controlled.}

Superscripts $\mathrm{uc}$, $\mathrm{sc}$, $\mathrm{host}$, $\mathrm{def}$, and $\mathrm{bg}$ denote unit-cell, supercell, host, defect, and background quantities, respectively.
Subscripts indicate wave-vector domains, reduced coordinates, element or cell indices, and degree-of-freedom partitions.

\subsection{Finite-Element Formulation}

We consider a two-dimensional periodic elastic medium under in-plane motion.
The numerical examples use plane strain and unit out-of-plane thickness.
After finite-element discretization, the undamped free-vibration problem for an assembled domain is
\begin{equation}
  \left(\boldsymbol{K} - \omega_i^2 \boldsymbol{M}\right)\boldsymbol{u}_i = \boldsymbol{0},
  \label{eq:full_evp}
\end{equation}
where $\boldsymbol{K}$ and $\boldsymbol{M}$ are the global stiffness and mass matrices, respectively.
The quantity $\omega_i$ is the $i$th angular eigenfrequency, and $\boldsymbol{u}_i$ is the corresponding displacement eigenvector.
The cyclic frequency is $f_i=\omega_i/(2\pi)$. Frequencies and targets are reported in Hz; $\omega_i$ is retained in eigenproblems and eigenfrequency sensitivities.

\subsection{Bloch Unit-Cell and Supercell Analysis}

For a periodic unit cell with lattice vectors $\boldsymbol{a}_1$ and $\boldsymbol{a}_2$, Bloch-Floquet conditions relate displacement vectors on opposite boundaries:
\begin{equation}
  \boldsymbol{u}(\boldsymbol{x}+\boldsymbol{a}_m)
  = \boldsymbol{u}(\boldsymbol{x})e^{\mathrm{i}\boldsymbol{k}\cdot \boldsymbol{a}_m},
  \qquad m=1,2,
  \label{eq:bloch}
\end{equation}
where $\boldsymbol{x}$ is the position vector and $\boldsymbol{k}$ is the wave vector.
The symbol $\mathrm{i}$ denotes the imaginary unit.
Applying the Bloch--Floquet condition in Eq.~\eqref{eq:bloch} yields the wave-vector-dependent eigenproblem
\begin{equation}
  \left[\boldsymbol{K}(\boldsymbol{k})-\omega_i^2(\boldsymbol{k})\boldsymbol{M}(\boldsymbol{k})\right]\boldsymbol{q}_i(\boldsymbol{k})=\boldsymbol{0}.
  \label{eq:bloch_evp}
\end{equation}\par

\begin{figure}[!tb]
\centering
\begin{subfigure}[t]{0.4\textwidth}
\centering
\includegraphics[width=\linewidth,trim=0 5.63bp 0 0,clip]{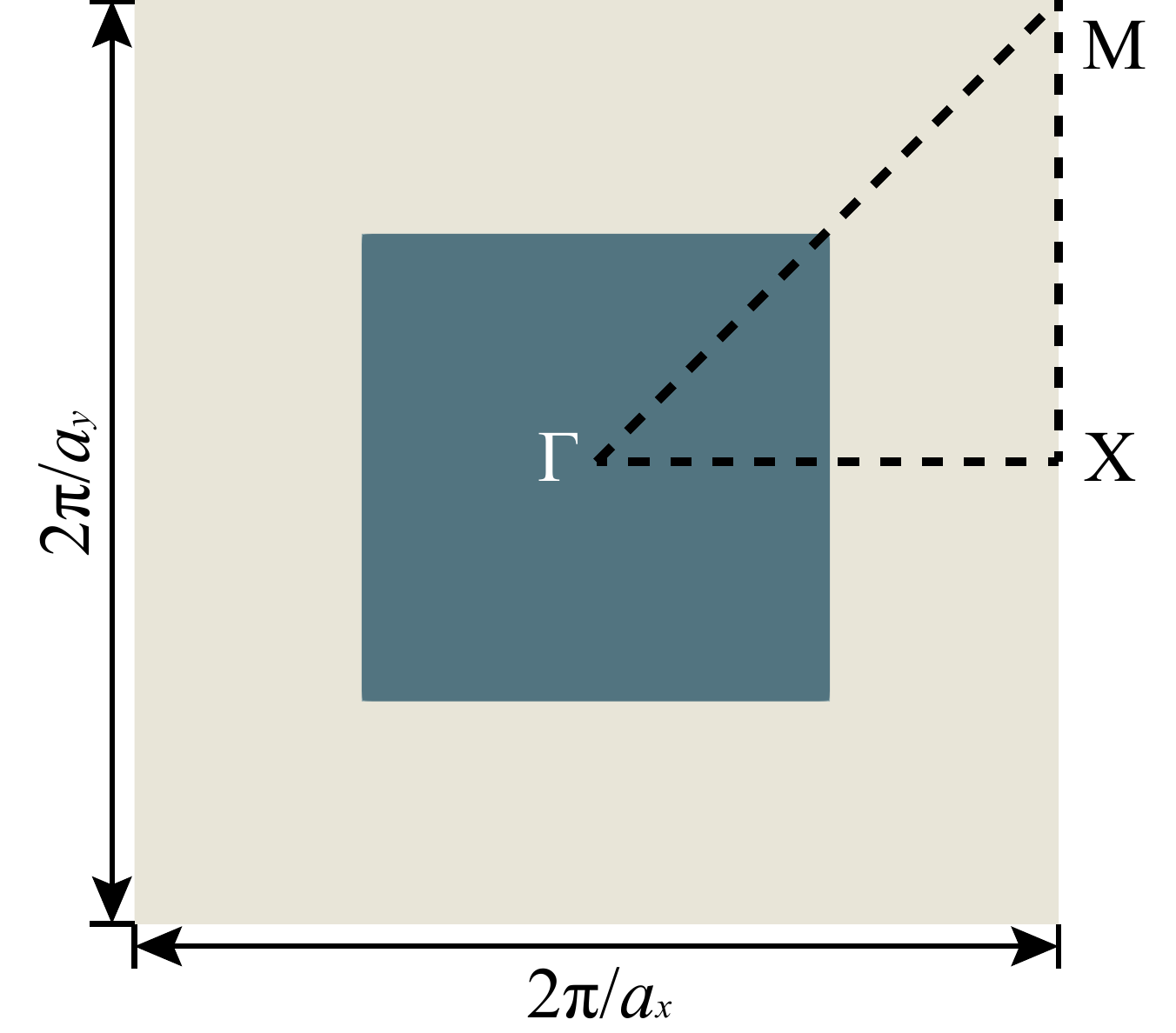}
\panelcaption{Unit-cell Brillouin zone.}
\end{subfigure}
\hfill
\begin{subfigure}[t]{0.4\textwidth}
\centering
\includegraphics[width=\linewidth,trim=0 2.49bp 0 0,clip]{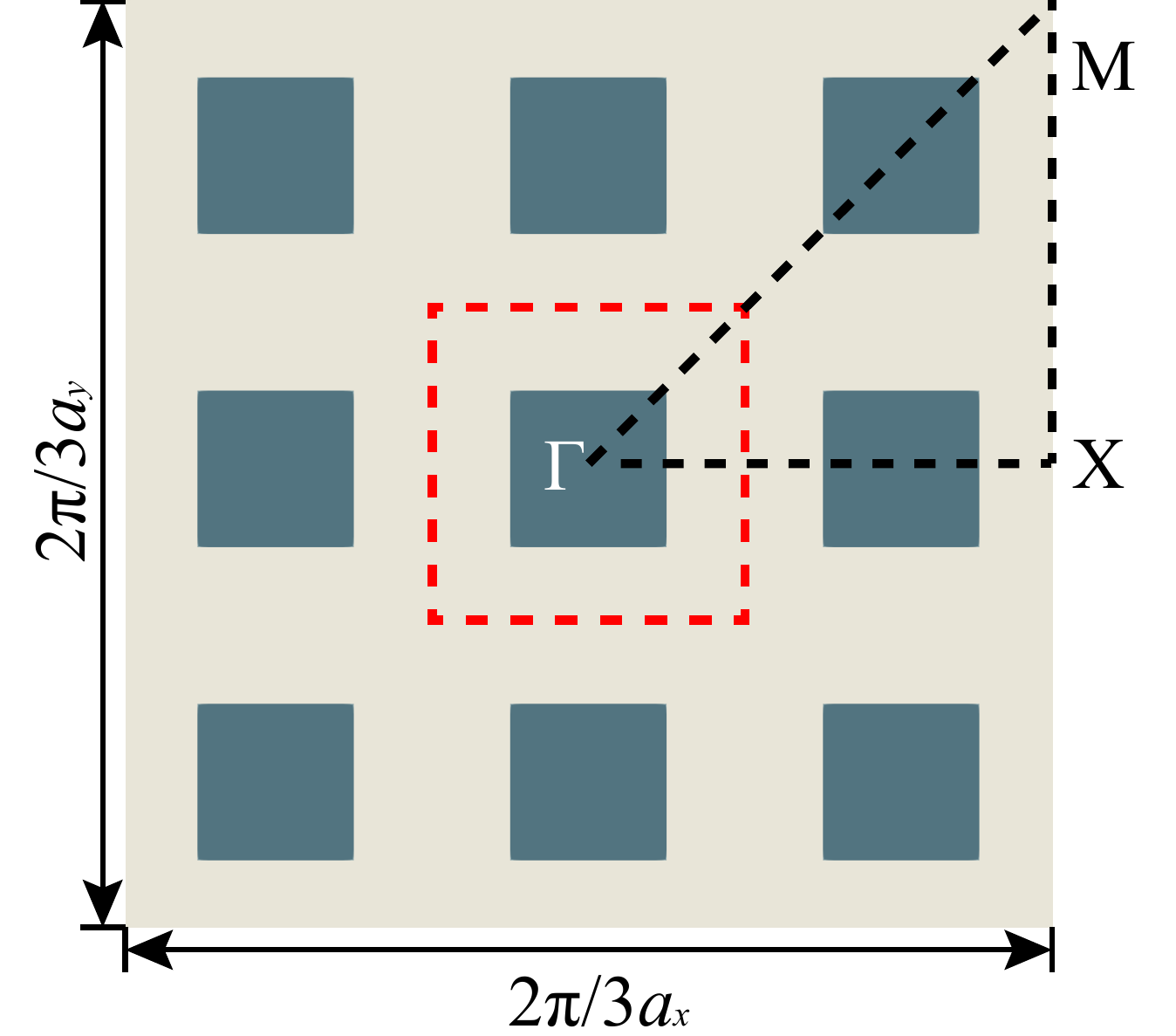}
\panelcaption{$3\times3$ supercell Brillouin zone.}
\end{subfigure}
\caption{Unit-cell and $3\times3$ supercell Brillouin zones with the sampled $\Gamma$--$\mathrm{X}$--$\mathrm{M}$--$\Gamma$ paths.}
\label{fig:ibz_paths}
\end{figure}

Here $\boldsymbol{K}(\boldsymbol{k})$ and $\boldsymbol{M}(\boldsymbol{k})$ are the Bloch-reduced stiffness and mass matrices, and $\boldsymbol{q}_i(\boldsymbol{k})$ is the Bloch eigenvector associated with the $i$th branch.
The dispersion relation is obtained by solving Eq.~\eqref{eq:bloch_evp} along the boundary of the irreducible Brillouin zone.
In the numerical implementation, a bandgap is identified over the sampled irreducible Brillouin zone path when no sampled branch intersects the interval.
For the orthogonal cells in Figure~\ref{fig:ibz_paths}, the unit-cell Brillouin zone has dimensions $2\pi/a_x$ by $2\pi/a_y$, where $a_x=\|\boldsymbol{a}_1\|$ and $a_y=\|\boldsymbol{a}_2\|$.
The $3\times3$ supercell dimensions are $2\pi/(3a_x)$ by $2\pi/(3a_y)$; both dispersions are sampled along the $\Gamma$--$\mathrm{X}$--$\mathrm{M}$--$\Gamma$ path.

A defect is modeled by replacing the central cell of an $N_x \times N_y$ supercell while keeping the surrounding cells identical to the host unit cell.
Here $N_x$ and $N_y$ are the numbers of cells in the two lattice directions.
Bloch periodicity is then imposed on the supercell boundary.
The supercell has larger lattice vectors and a correspondingly folded Brillouin zone.
Its eigenproblem is
\begin{equation}
  \left[\boldsymbol{K}^{\mathrm{sc}}(\boldsymbol{k}_{\mathrm{sc}})
  - \omega_i^2(\boldsymbol{k}_{\mathrm{sc}})
  \boldsymbol{M}^{\mathrm{sc}}(\boldsymbol{k}_{\mathrm{sc}})\right]
  \boldsymbol{q}_i^{\mathrm{sc}}(\boldsymbol{k}_{\mathrm{sc}})=\boldsymbol{0}.
  \label{eq:supercell_bloch}
\end{equation}\par

Here $\boldsymbol{K}^{\mathrm{sc}}(\boldsymbol{k}_{\mathrm{sc}})$ and $\boldsymbol{M}^{\mathrm{sc}}(\boldsymbol{k}_{\mathrm{sc}})$ are the Bloch-reduced supercell stiffness and mass matrices.
The vector $\boldsymbol{k}_{\mathrm{sc}}$ is the supercell wave vector, and $\boldsymbol{q}_i^{\mathrm{sc}}$ is the corresponding supercell eigenvector.
In the absence of a defect, the supercell dispersion is a folded representation of the unit-cell dispersion.
When the central cell is modified, additional nearly flat branches may appear inside the host bandgap.
Such nearly flat in-gap branches are referred to as defect bands in this study.
\rev{In the supercell dispersion plots, the defect-band window denotes the shaded interval enclosing the defect bands between the indicated lower and upper bounding branches along the sampled wave-vector path. This window depends on the defect design and is distinct from the fixed Stage~I host bandgap. It contains allowed defect bands, including competing bands in the attraction-only comparison; target-to-non-target separation is evaluated separately.}

\section{\rev{Spectrum-selective optimization and reduced assembly}}
\label{sec:optimization}

Stage~I establishes a host bandgap containing the prescribed operating frequency.
Stage~II optimizes the central defect cell to control the current $\Gamma$-point in-gap spectrum, using reduced-order analysis to limit the cost of repeated supercell eigensolves; localization is evaluated afterward.

\subsection{Stage I: Target-Centered Host Bandgap Design}

Stage I optimizes the periodic host unit cell to create a host bandgap identified along the sampled irreducible Brillouin-zone path and containing the prescribed bandgap-center frequency $f^*$.
The formulation follows the target-centered bandgap method of Xu et al. \cite{xuTwostageTopologyOptimization2026}.

The Stage~I optimization problem is
\begin{equation}
  \begin{aligned}
  \max_{\boldsymbol{s}^{\mathrm{uc}}}\quad
  & J^{\mathrm{uc}}(\boldsymbol{s}^{\mathrm{uc}})
  =L_{\mathrm{M}}(\boldsymbol{s}^{\mathrm{uc}}),\\
  \mathrm{s.t.}\quad
  & g^{\mathrm{uc}}(\boldsymbol{s}^{\mathrm{uc}})
  =\frac{V(\boldsymbol{s}^{\mathrm{uc}})}{V_{\mathrm{lim}}^{\mathrm{uc}}}-1\leq0,\\
  & h^{\mathrm{uc}}(\boldsymbol{s}^{\mathrm{uc}})\leq0,\\
  & 0<s_{\min}\leq s_e^{\mathrm{uc}}\leq1,
  \qquad e=1,\ldots,N_e^{\mathrm{uc}}.
  \end{aligned}
  \label{eq:stage1}
\end{equation}\par

Here $\boldsymbol{s}^{\mathrm{uc}}$ is the unit-cell design vector, $s_e^{\mathrm{uc}}$ is its $e$-th element variable, and $N_e^{\mathrm{uc}}$ is the number of unit-cell design elements.
The function $V(\boldsymbol{s}^{\mathrm{uc}})$ gives the current material volume, $V_{\mathrm{lim}}^{\mathrm{uc}}$ specifies its upper limit, and $s_{\min}$ is the common lower bound on the element variables.
The objective $L_{\mathrm{M}}$ is a normalized Kreisselmeier--Steinhauser aggregation of the squared distances between the relevant band extrema and $f^*$.
The constraint $h^{\mathrm{uc}}$ is the quadratic exclusion aggregation that prevents dispersion branches from crossing $f^*$, whereas $g^{\mathrm{uc}}$ imposes the material-volume limit.
The detailed aggregation, crossing constraint, and sensitivity expressions are given by Xu et al.~\cite{xuTwostageTopologyOptimization2026}.
After convergence, the optimized host topology and the resulting lower and upper bandgap bounds $[f_{\mathrm{L}},f_{\mathrm{U}}]$ are passed to Stage~II.

\subsection{Stage II: Spectrum-Selective Defect-Mode Control within Fixed Host-Bandgap Bounds}

Stage~II optimizes only the central defect cell while keeping the optimized host cells fixed.
The Stage~II target frequency $f^{**}$ is selected within the host bandgap $[f_{\mathrm{L}},f_{\mathrm{U}}]$ established in Stage~I.
It may differ from the Stage~I bandgap-center frequency $f^*$, so its offset from the resulting host-bandgap midpoint is retained explicitly in the Stage~II objective.
The design vector $\boldsymbol{s}^{\mathrm{def}}$ therefore contains only the central-cell element variables.

The objective attracts selected in-gap modes toward $f^{**}$ and repels the remaining in-gap modes from the host-bandgap midpoint. \rev{It is evaluated at $\Gamma$; dispersion and localization are assessed after optimization.}
The complete Stage II problem is
\begin{equation}
  \begin{aligned}
  \min_{\boldsymbol{s}^{\mathrm{def}}}\quad
  & {
  J^{\mathrm{def}}(\boldsymbol{s}^{\mathrm{def}})
  =
  \sum_{j\in\mathcal{D}}
  \left[
  S(f_j) P_{\mathrm{a}}(f_j)
  + \lambda\left(1-S(f_j)\right) P_{\mathrm{r}}(f_j)
  \right]},\\
  \mathrm{s.t.}\quad
  & g^{\mathrm{def}}(\boldsymbol{s}^{\mathrm{def}})=
  \frac{V(\boldsymbol{s}^{\mathrm{def}})}{V_{\mathrm{lim}}^{\mathrm{def}}}-1\leq 0,\\
  & 0<s_{\min}\leq s_e^{\mathrm{def}}\leq 1,\qquad e=1,\ldots,N_e^{\mathrm{def}} .
  \end{aligned}
  \label{eq:stage2}
\end{equation}\par

Here $\mathcal{D}=\{j\,|\,f_{\mathrm{L}}\leq f_j\leq f_{\mathrm{U}}\}$ is the current set of $\Gamma$-point in-gap modes. It is re-identified after each eigensolve. \rev{The selection weights vary smoothly with frequency; the objective is treated as piecewise smooth between changes in in-gap membership.}
The normalized constraint $g^{\mathrm{def}}$ limits the defect-cell volume $V$ to $V_{\mathrm{lim}}^{\mathrm{def}}$; $N_e^{\mathrm{def}}$ is the number of variables, each bounded below by $s_{\min}$.
The parameter $\lambda$ balances the attraction and repulsion terms weighted by $S(f_j)$.
The attraction penalty $P_{\mathrm{a}}$ is defined as
\begin{equation}
  {
  P_{\mathrm{a}}(f_j)=
  \left(\frac{f_j-f^{**}}{f^{**}}\right)^2.}
  \label{eq:attraction_ijms}
\end{equation}\par

\begin{figure}[!b]
\centering
\begin{subfigure}[t]{0.46\textwidth}
\includegraphics[width=\linewidth,trim=0 12.00bp 0 15.00bp,clip]{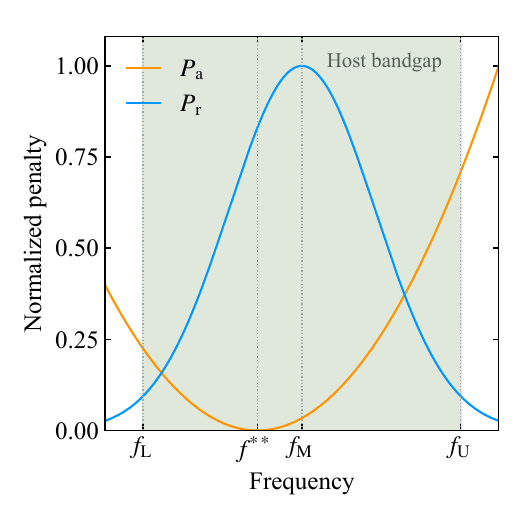}
\panelcaption{Attraction and repulsion.}
\end{subfigure}\hfill
\begin{subfigure}[t]{0.46\textwidth}
\includegraphics[width=\linewidth,trim=0 13.00bp 0 15.00bp,clip]{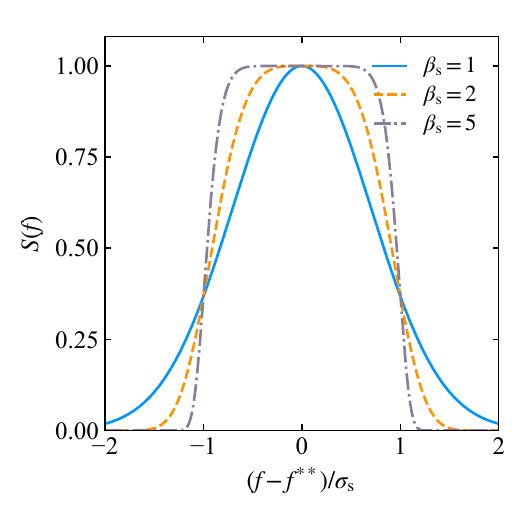}
\panelcaption{Selection function.}
\end{subfigure}
\caption{Spectral weighting in Stage~II. \rev{In (a), attraction is centered at the target frequency and repulsion at the host-bandgap midpoint; the two penalties are scaled independently for illustration. In (b), frequency is normalized by the selection width. Increasing $\beta_{\mathrm{s}}$ flattens the selected neighborhood and sharpens its edges, as prescribed by Eq.~\eqref{eq:selection}.}}
\label{fig:stage2_objective_terms}
\label{fig:stage2_selection_function}
\end{figure}

This term penalizes deviation from the prescribed defect frequency.
The repulsion penalty $P_{\mathrm{r}}$ is centered at the host bandgap midpoint $f_{\mathrm{M}}=(f_{\mathrm{L}}+f_{\mathrm{U}})/2$:
 \begin{equation}
  { P_{\mathrm{r}}(f_j)=
  \exp\left[
  -\frac{(f_j-f_{\mathrm{M}})^2}{2\sigma_{\mathrm{r}}^2}
  \right],}
  \label{eq:repulsion_ijms}
\end{equation}
where $\sigma_{\mathrm{r}}$ is the repulsion width in Hz.
Figure~\ref{fig:stage2_objective_terms} contrasts the two centers: the target-selection function defined below is centered on $f^{**}$, whereas $P_{\mathrm{r}}$ is centered on $f_{\mathrm{M}}$. When $f^{**}\neq f_{\mathrm{M}}$, the selection function assigns target-adjacent modes predominantly to the attraction component, whereas minimizing the repulsion component drives unselected in-gap frequencies away from the host-bandgap midpoint.

The selection function is defined as
\begin{equation}
  {
  S(f_j)=
  e^{-\left(\frac{f_j-f^{**}}{\sigma_{\mathrm{s}}}\right)^{2\beta_{\mathrm{s}}}},}
  \label{eq:selection}
\end{equation}\par

Here $\sigma_{\mathrm{s}}$ controls the selection width, and the integer $\beta_{\mathrm{s}}\geq1$ controls transition sharpness while retaining a real, symmetric function.
As illustrated in Figure~\ref{fig:stage2_selection_function}, Eq.~\eqref{eq:selection} is Gaussian for $\beta_{\mathrm{s}}=1$; larger values flatten the target neighborhood and sharpen the transition. The numerical examples use $\beta_{\mathrm{s}}=5$.

\rev{Nearest-neighbor separation is evaluated from the optimized spectrum; the repulsion penalty itself is defined relative to the host-bandgap midpoint.}

\subsection{\rev{Reduced assembly with local defect updates}}
\label{sec:assembly}

Stage~II requires repeated eigenvalue solutions for large supercells.
Increasing the supercell size raises both the number of displacement degrees of freedom and the number of folded branches that must be computed over a given frequency range.
A $3\times3$ supercell has approximately nine times the unit-cell degrees of freedom, and band folding produces approximately nine times as many branches in the same interval.
The resulting cost increase is especially pronounced when many modes are requested, because eigensolver orthogonalization and projection costs also grow with the subspace dimension.
A Craig--Bampton (CB) formulation reduces this cost while retaining the interface coordinates that transmit defect--host coupling.

\rev{Since Stage~II changes only the central cell, the reduced host contribution can be reused throughout the optimization.}

\subsubsection{Background-Defect Decomposition}

\rev{To separate the reusable and evolving matrix contributions, the reduced supercell is partitioned into the background cells and the central defect cell.} An overbar denotes CB-reduced coordinates; $\mathrm{host}$ identifies the optimized host cell, whereas $\mathrm{bg}$ is the assembled contribution of all non-defect cells.
For $N_{\mathrm{cell}}=N_xN_y$ cells, let $c_{\mathrm{d}}$ be the defect-cell index and $\mathcal{C}_{\mathrm{bg}}=\{1,\ldots,N_{\mathrm{cell}}\}\setminus\{c_{\mathrm{d}}\}$ the background-cell set.
The reduced localization matrix $\boldsymbol{\Lambda}_c$ maps global reduced supercell coordinates to the reduced coordinates of cell $c$:
\begin{equation}
    \boldsymbol{z}_c
    =
    \boldsymbol{\Lambda}_c\boldsymbol{z}_{\mathrm{sc}},
    \qquad
    \boldsymbol{\Lambda}_c\in\mathbb{R}^{m_c\times m_{\mathrm{sc}}},
    \label{eq:reduced_localization}
\end{equation}
 where $\boldsymbol{z}_{\mathrm{sc}}$ is the global reduced supercell vector and $\boldsymbol{z}_c$ is the reduced coordinate vector of cell $c$.
The cell-level reduced dimension is $m_c=n_{\mathrm{fi}}+n_{\mathrm{b}}$, where $n_{\mathrm{fi}}$ is the number of retained fixed-interface modes and $n_{\mathrm{b}}$ is the number of physical boundary degrees of freedom retained for one cell.
At optimization iteration $n$, the CB-reduced supercell matrices are expressed as
  \begin{equation}
    \bar{\boldsymbol{K}}^{\mathrm{sc},[n]}
    =
    \bar{\boldsymbol{K}}^{\mathrm{bg}}
    +
    \boldsymbol{\Lambda}_{c_{\mathrm{d}}}^{\mathrm{T}}
    \bar{\boldsymbol{K}}^{\mathrm{def},[n]}
    \boldsymbol{\Lambda}_{c_{\mathrm{d}}},
    \quad
    \bar{\boldsymbol{M}}^{\mathrm{sc},[n]}
    =
    \bar{\boldsymbol{M}}^{\mathrm{bg}}
    +
    \boldsymbol{\Lambda}_{c_{\mathrm{d}}}^{\mathrm{T}}
    \bar{\boldsymbol{M}}^{\mathrm{def},[n]}
    \boldsymbol{\Lambda}_{c_{\mathrm{d}}}.
    \label{eq:supercell_iterative_update}
\end{equation}\par

The superscript $[n]$ denotes the optimization iteration; $\bar{\boldsymbol{K}}^{\mathrm{bg}}$ and $\bar{\boldsymbol{M}}^{\mathrm{bg}}$ are constant, whereas the defect-cell contributions depend on $n$.

The subscripts $\mathrm{i}$ and $\mathrm{b}$ denote interior and boundary degrees of freedom, respectively.
\rev{For either cell type $\alpha\in\{\mathrm{host},\mathrm{def}\}$, the retained fixed-interface modes and constraint modes satisfy}

\begin{align}
\boldsymbol{K}_{\mathrm{ii}}^\alpha\boldsymbol{\Phi}_{\mathrm{fi}}^\alpha
&=\boldsymbol{M}_{\mathrm{ii}}^\alpha\boldsymbol{\Phi}_{\mathrm{fi}}^\alpha\boldsymbol{\Omega}_{\mathrm{fi}}^\alpha,
\label{eq:host_fi_modes}\\
\boldsymbol{K}_{\mathrm{ii}}^\alpha\boldsymbol{\Psi}_{\mathrm{ib}}^\alpha
&=-\boldsymbol{K}_{\mathrm{ib}}^\alpha.
\label{eq:host_constraint_modes}
\end{align}

\rev{Here $\boldsymbol{\Omega}_{\mathrm{fi}}^\alpha$ contains the retained squared eigenfrequencies.} With physical coordinates ordered as interior and boundary displacements, the cell projection is
\begin{equation}
\begin{aligned}
\boldsymbol{T}_{\mathrm{CB}}^{\alpha}
&=\begin{bmatrix}
\boldsymbol{\Phi}_{\mathrm{fi}}^{\alpha} & \boldsymbol{\Psi}_{\mathrm{ib}}^{\alpha}\\
\boldsymbol{0} & \boldsymbol{I}
\end{bmatrix},\\
\bar{\boldsymbol{K}}^{\alpha}
&=(\boldsymbol{T}_{\mathrm{CB}}^{\alpha})^{\mathrm{T}}\boldsymbol{K}^{\alpha}\boldsymbol{T}_{\mathrm{CB}}^{\alpha},
\qquad
\bar{\boldsymbol{M}}^{\alpha}
=(\boldsymbol{T}_{\mathrm{CB}}^{\alpha})^{\mathrm{T}}\boldsymbol{M}^{\alpha}\boldsymbol{T}_{\mathrm{CB}}^{\alpha}.
\end{aligned}
\label{eq:cell_cb_projection}
\end{equation}

The reduced coordinates are ordered as retained fixed-interface modal amplitudes followed by physical boundary displacements.
The fixed-interface modes approximate the cell interior dynamics, while the retained physical boundary coordinates and constraint modes transmit displacement compatibility between adjacent cells.
The precomputed background contribution is
  \begin{equation}
    \bar{\boldsymbol{K}}^{\mathrm{bg}}
    =
    \sum_{c\in\mathcal{C}_{\mathrm{bg}}}
    \boldsymbol{\Lambda}_c^{\mathrm{T}}
    \bar{\boldsymbol{K}}^{\mathrm{host}}
    \boldsymbol{\Lambda}_c,
    \quad
    \bar{\boldsymbol{M}}^{\mathrm{bg}}
    =
    \sum_{c\in\mathcal{C}_{\mathrm{bg}}}
    \boldsymbol{\Lambda}_c^{\mathrm{T}}
    \bar{\boldsymbol{M}}^{\mathrm{host}}
    \boldsymbol{\Lambda}_c.
    \label{eq:background_precomputation}
\end{equation}\par

\rev{Because the physical interface coordinates remain unchanged, the updated defect basis is compatible with the fixed host bases and uses the same localization matrix. Host-basis construction and background assembly are performed once. At each iteration, the defect basis and reduced matrices are recomputed and the coupled reduced supercell eigenproblem is solved. For a $5\times5$ supercell, the contributions of all 24 host cells are reused.}

\subsubsection{Reduced-Order Assembly}

After shared interfaces are merged, the reduced supercell dimension is
\begin{equation}
    m_{\mathrm{sc}}
    =
    N_{\mathrm{cell}} n_{\mathrm{fi}}
    +
    n_{\mathrm{b,global}},
    \label{eq:supercell_reduced_dimension}
\end{equation}
where $n_{\mathrm{b,global}}$ is the number of unique boundary degrees of freedom after shared interfaces are merged.

\rev{Interface reduction can further decrease the system size when boundary coordinates dominate, as studied for enhanced CB models by Bondsman et al.~\cite{bondsmanInterfaceReductionTechnique2024}. Here retaining physical interface coordinates permits local basis updates, but the global solve retains the interface-size contribution in Eq.~\eqref{eq:supercell_reduced_dimension}.}
The matrices $\bar{\boldsymbol{K}}^{\mathrm{sc},[n]}(\boldsymbol{k}_{\mathrm{sc}})$ and $\bar{\boldsymbol{M}}^{\mathrm{sc},[n]}(\boldsymbol{k}_{\mathrm{sc}})$ are obtained from Eq.~\eqref{eq:supercell_iterative_update} by imposing the supercell Bloch condition on the outer boundary.
The CB-reduced supercell eigenproblem is
\begin{equation}
    \left[
    \bar{\boldsymbol{K}}^{\mathrm{sc},[n]}(\boldsymbol{k}_{\mathrm{sc}})
    -
    \omega_i^2(\boldsymbol{k}_{\mathrm{sc}})
    \bar{\boldsymbol{M}}^{\mathrm{sc},[n]}(\boldsymbol{k}_{\mathrm{sc}})
    \right]
    \bar{\boldsymbol{q}}_i^{\mathrm{sc}}(\boldsymbol{k}_{\mathrm{sc}})
    =
    \boldsymbol{0}.
    \label{eq:reduced_supercell_bloch}
\end{equation}\par

Here $\bar{\boldsymbol{q}}_i^{\mathrm{sc}}$ is the reduced supercell eigenvector; during Stage~II, Eq.~\eqref{eq:reduced_supercell_bloch} is evaluated at $\boldsymbol{k}_{\mathrm{sc}}=\boldsymbol{0}$.
After optimization, the same reduced assembly can be evaluated along the supercell Brillouin-zone path to compute the reduced-order dispersion relation.

\subsection{Optimization Workflow}

\rev{Algorithm~\ref{alg:merged_workflow} combines the spectral objective and reduced assembly in the two-stage optimization. During Stage~II, narrowing the selection width concentrates attraction around the target, while increasing the repulsion weight strengthens non-target repulsion. Table~\ref{tab:spectral_parameters} gives these common parameter schedules.}

\begin{table}[!b]
\centering

\caption{\rev{Common Stage~II spectral parameters. The index $n$ counts design updates; widths are expressed in Hz.}}
\label{tab:spectral_parameters}
\small
\begin{tabular}{@{}p{0.23\textwidth}p{0.70\textwidth}@{}}
\toprule
\rev{Quantity} & \rev{Rule and parameter values} \\
\midrule
\rev{Repulsion width} & \rev{$\sigma_{\mathrm r}=0.3(f_{\mathrm U}-f_{\mathrm L})$} \\[4pt]
\rev{Selection width} & \rev{$\sigma_{\mathrm s}^{[n]}=f^{**}\max[0.02,\,0.5e^{-0.05\max(0,n-50)}]$} \\[4pt]
\rev{Repulsion weight} & \rev{$\lambda^{[n]}=\min[1.0,\,0.1e^{0.05\max(0,n-5)}]$} \\[4pt]
\rev{Selection exponent} & \rev{$\beta_{\mathrm s}=5$} \\
\bottomrule
\end{tabular}
\end{table}

\begin{algorithm}[!htb]
\caption{\rev{Spectrum-selective optimization with a reusable reduced background.}}
\label{alg:merged_workflow}
\begingroup\renewcommand{\baselinestretch}{1}\small\selectfont
\begin{algorithmic}[1]
\Require \rev{Host and defect targets, volume limits, supercell size, retained mode count}
\State \rev{Optimize the host using Eq.~\eqref{eq:stage1}; fix its topology and bandgap bounds.}
\State \rev{Construct the host CB basis and assemble the fixed background contribution.}
\State \rev{Initialize the central defect design and continuation parameters.}
\While{\rev{the stopping criterion is not met}}
  \State \rev{Filter the defect variables and evaluate the physical material properties.}
  \State \rev{Update the defect CB basis and its reduced stiffness and mass matrices.}
  \State \rev{Insert the defect contribution into the precomputed background.}
  \State \rev{Solve the reduced $\Gamma$-point eigenproblem and identify the in-gap set.}
  \State \rev{Evaluate the selection-weighted objective in Eq.~\eqref{eq:stage2}.}
  \State \rev{Recover modal fields and evaluate the frozen-basis sensitivities.}
  \State \rev{Apply the filter chain rule and update the design with MMA.}
  \State \rev{Update the spectral weights using Table~\ref{tab:spectral_parameters} and the filter radius using \ref{sec:implementation}.}
\EndWhile
\State \rev{Reanalyze the final design with the full-order model at $\Gamma$.}
\State \rev{Evaluate the target spectrum, cluster localization, and dispersion.}
\end{algorithmic}
\endgroup
\end{algorithm}

\rev{In both stages, convergence is declared when the change in the design variables $\boldsymbol{s}$ remains below $10^{-3}$ for five consecutive design updates. The same criterion is used in the matched full- and reduced-order runs.}

The method of moving asymptotes (MMA) \cite{svanbergMethodMovingAsymptotes1987} is used for the design-variable updates.

\section{\rev{Mode recovery and sensitivity evaluation}}
\label{sec:sensitivity}

\rev{Updating the defect design requires derivatives of the spectral objective with respect to the element variables.} For Stage~II, reduced eigenvectors are recovered to full-order displacement fields to evaluate the elementwise stiffness and mass contributions. Stage~I sensitivities follow the target-centered bandgap implementation in Ref.~\cite{xuTwostageTopologyOptimization2026}.
\rev{The filtered design variable is the material fraction; the filtering, RAMP interpolation, and their derivatives are given in \ref{sec:implementation}.} For readability, the iteration superscript $[n]$ is omitted below.

\subsection{Mode Recovery and Eigenfrequency Sensitivity}

Because material derivatives use full-order element matrices, each reduced $\Gamma$-point eigenvector is recovered as
\begin{equation}
  \boldsymbol{u}_i \approx \boldsymbol{T}_{\mathrm{sc}}\bar{\boldsymbol{q}}_i^{\mathrm{sc}}(\boldsymbol{0}).
  \label{eq:mode_recovery}
\end{equation}\par

Here $\boldsymbol{T}_{\mathrm{sc}}$, assembled from the cell-level CB transformations and localization matrices, maps reduced supercell coordinates to the full-order displacement field.

The defect-cell CB basis is updated at each iteration and held fixed during sensitivity evaluation. \rev{The resulting frozen-basis derivative neglects the design dependence of the basis; its approximation to the full-order eigenfrequency derivative is assessed for the configuration in Section~\ref{sec:gradient_check}.}

For a simple reduced eigenvalue with the basis held fixed, the eigenfrequency derivative with respect to a filtered design variable is
\begin{equation}
  \frac{\partial \omega_i}{\partial \tilde{s}_e}
  =
  \frac{1}{2\omega_i}
  \frac{\boldsymbol{u}_i^{\mathrm{T}}
  \left(
  \dfrac{\partial \boldsymbol{K}}{\partial \tilde{s}_e}
  - \omega_i^2 \dfrac{\partial \boldsymbol{M}}{\partial \tilde{s}_e}
  \right)
  \boldsymbol{u}_i}{\boldsymbol{u}_i^{\mathrm{T}} \boldsymbol{M}\boldsymbol{u}_i}.
  \label{eq:eig_sens}
\end{equation}\par

In Eq.~\eqref{eq:eig_sens}, only elements in the central defect cell contribute to the full-order matrix derivatives.
For the cyclic-frequency objective, $\partial f_i/\partial{\tilde{s}_e}=(2\pi)^{-1}\partial\omega_i/\partial{\tilde{s}_e}$.

For an exactly repeated squared eigenfrequency $\mu=\omega^2$, let $\mathcal{C}$ denote the complete repeated cluster and let $\boldsymbol{U}_{\mathcal{C}}=[\boldsymbol{u}_a]_{a\in\mathcal{C}}$ contain an $\boldsymbol{M}$-orthonormal basis, such that $\boldsymbol{U}_{\mathcal{C}}^{\mathrm{T}}\boldsymbol{M}\boldsymbol{U}_{\mathcal{C}}=\boldsymbol{I}$.
Following the generalized gradient-matrix treatment of Seyranian et al.~\cite{seyranianMultipleEigenvaluesStructural1994}, the matrix for element variable $\tilde{s}_e$ is
\begin{equation}
  G_{ab}^{(e)}
  =
  \boldsymbol{u}_a^{\mathrm{T}}
  \left(
  \frac{\partial \boldsymbol{K}}{\partial \tilde{s}_e}
  -
  \mu
  \frac{\partial \boldsymbol{M}}{\partial \tilde{s}_e}
  \right)
  \boldsymbol{u}_b,
  \qquad a,b\in\mathcal{C}.
  \label{eq:multiple_eigen_gradient_matrix}
\end{equation}\par

The individual directional derivatives of the repeated squared eigenfrequencies are the eigenvalues of $\boldsymbol{G}^{(e)}$ and need not be assigned to fixed modal labels.
For the present symmetric spectral objective, let $\varphi(f)$ denote the single-mode contribution in Eq.~\eqref{eq:stage2}.  All members of an exact repeated cluster have the same $f$ and hence the same $\varphi'(f)$.  The derivative of the complete cluster contribution $J_{\mathcal{C}}=\sum_{a\in\mathcal{C}}\varphi(f_a)$ is therefore
\begin{equation}
  {
  \frac{\partial J_{\mathcal{C}}}{\partial\tilde{s}_e}
  =
  \frac{\varphi'(f)}{4\pi\omega}
  \operatorname{tr}\!\left(\boldsymbol{G}^{(e)}\right).}
  \label{eq:repeated_cluster_objective_gradient}
\end{equation}\par

Because the trace equals the sum of the diagonal Rayleigh terms in any $\boldsymbol{M}$-orthonormal basis, Eq.~\eqref{eq:repeated_cluster_objective_gradient} is exactly the basis-invariant cluster contribution obtained by the per-mode accumulation in Eq.~\eqref{eq:stage2_sens_chain}.  Distinct but closely spaced eigenvalues are treated with the simple-eigenvalue expression in Eq.~\eqref{eq:eig_sens}.

\subsection{Stage~II Objective Sensitivity}

\rev{The eigenfrequency derivatives are combined with the attraction, repulsion, and selection-weight derivatives to obtain the Stage~II objective gradient.} For the current in-gap set $\mathcal{D}$ in Eq.~\eqref{eq:stage2}, the derivative with respect to a filtered defect design variable is
\begin{equation}
  {
  \frac{\partial J^{\mathrm{def}}}{\partial \tilde{s}_e}
  =
  \sum_{j\in\mathcal{D}}
  \left[
  \frac{\partial J^{\mathrm{def}}}{\partial f_j}
  \frac{\partial f_j}{\partial \tilde{s}_e}
  \right].}
  \label{eq:stage2_sens_chain}
\end{equation}\par

The scalar derivative with respect to an in-gap frequency is
\begin{equation}
  {
  \frac{\partial J^{\mathrm{def}}}{\partial f_j}
  =
  S(f_j)
  \frac{\partial P_{\mathrm{a}}}{\partial f_j}
  +\lambda\left[1-S(f_j)\right]
  \frac{\partial P_{\mathrm{r}}}{\partial f_j}
  +
  \frac{\partial S}{\partial f_j}
  \left[
  P_{\mathrm{a}}(f_j)-\lambda P_{\mathrm{r}}(f_j)
  \right].}
  \label{eq:stage2_total_freq_derivative}
\end{equation}\par

The three terms in Eq.~\eqref{eq:stage2_total_freq_derivative} account for the attraction penalty, the repulsion penalty, and the change in their selection weight as a mode moves through the target neighborhood.

The component derivatives are
 \begin{align}
  {\frac{\partial P_{\mathrm{a}}}{\partial f_j}
  }&{=
  \frac{2(f_j-f^{**})}{(f^{**})^2},}\\
  {\frac{\partial P_{\mathrm{r}}}{\partial f_j}
  }&{=
  -\frac{f_j-f_{\mathrm{M}}}{\sigma_{\mathrm{r}}^2}
  P_{\mathrm{r}}(f_j),}\\
  {\frac{\partial S}{\partial f_j}
  }&{=
  -2\beta_{\mathrm{s}}
  \frac{(f_j-f^{**})^{2\beta_{\mathrm{s}}-1}}{\sigma_{\mathrm{s}}^{2\beta_{\mathrm{s}}}}
  S(f_j).}
  \label{eq:stage2_component_derivatives}
\end{align}\par

Finally, the derivative with respect to an unfiltered defect design variable follows from the filter chain rule:
\begin{equation}
  {
  \frac{\partial J^{\mathrm{def}}}{\partial s_\ell}
  =
  \sum_{e=1}^{N_e^{\mathrm{def}}}
  \frac{\partial J^{\mathrm{def}}}{\partial \tilde{s}_e}
  \frac{\partial \tilde{s}_e}{\partial s_\ell},
  }
  \qquad
  \frac{\partial g^{\mathrm{def}}}{\partial s_\ell}
  =
  \sum_{e=1}^{N_e^{\mathrm{def}}}
  \frac{\partial g^{\mathrm{def}}}{\partial \tilde{s}_e}
  \frac{\partial \tilde{s}_e}{\partial s_\ell}.
  \label{eq:filter_chain_rule_ijms}
\end{equation}\par

\section{Numerical Examples}
\label{sec:numerical}

The numerical study evaluates spectral and sensitivity fidelity, computational cost, and localized target-mode placement across different materials and supercell sizes, using dispersion post-processing, full-order $\Gamma$-point reanalysis, and a representative finite-array response.

\subsection{Problem Setup}

All cases use a square unit cell with side length $L=1~\mathrm{m}$ and report cyclic frequencies $f$ in Hz.
Tables~\ref{tab:material_configurations} and \ref{tab:case_configurations} list the material data, target frequencies, supercell sizes, and meshes; the 100 Hz case is also used for the finite-array response check.
The cases share the same basic discretization and volume setting while varying frequency range, material contrast, or supercell size, allowing these effects to be examined without changing the optimization formulation.

\begin{table}[!tb]
\centering
\caption{Material properties for the two-phase configurations.}
\label{tab:material_configurations}
\begin{tabular}{ccccc}
\toprule
  \multirow{2}{*}{\shortstack{Material\\configuration}}  & Material & Young's modulus & Density & Poisson's ratio \\
              & ID       & [GPa]           & [$\mathrm{kg/m}^3$] & [-] \\
\midrule
\multirow{2}{*}{A} & 1 & 0.02    & 1000  & 0.3  \\
                   & 2 & 0.5     & 3000 & 0.3  \\

\multirow{2}{*}{B} & 1 & 3    & 1180   & 0.35  \\
                   & 2 & 70     & 2700  & 0.35  \\
\bottomrule
\end{tabular}
\end{table}

Configuration A is used for the 100 Hz, 200 Hz, and $5 \times 5$ cases, whereas configuration B is used for the 1300 Hz case.

\begin{table}[!tb]
\centering
\caption{Case configurations used to evaluate material effects, target frequency, and supercell size.}
\label{tab:case_configurations}
\small
\setlength{\tabcolsep}{4pt}
\begin{tabular}{@{}ccccc@{}}
\toprule
\multirow{2}{*}{Case} & Material & Target & Supercell & Mesh density \\
 & configuration & frequency & size & (per unit cell) \\
\midrule
1 & A & 100 Hz & $3\times3$ & $60 \times 60$ \\
2 & A & 200 Hz & $3\times3$ & $60 \times 60$ \\
3 & B & 1300 Hz & $3\times3$ & $60 \times 60$ \\
4 & A & 100 Hz & $5\times5$ & $60 \times 60$ \\
\bottomrule
\end{tabular}
\end{table}

Mesh density denotes the number of bilinear quadrilateral (Q4) elements per unit cell, and both stages use a 50~\% volume limit.
Stage~I uses six Bloch branches for the lower-frequency searches and 18 for the 200 Hz search.
Dispersion plots use the corresponding optimization model order, except for the 200 Hz Stage~II case, where both optimized designs are reanalyzed with the full-order model along the complete plotted path. All localization measures use full-order $\Gamma$-point reanalysis.
All wall-clock timings were obtained with an OpenMP-enabled implementation on a computing server equipped with an Intel Xeon Gold 6136 CPU at 3.0 GHz and 192 GB of memory.
Speedups compare identical software and hardware settings.

\subsection{ Reduced-Order Representation of In-Gap Defect Modes}
\label{sec:reduced_spectrum_validation}

\begin{figure}[!b]
\centering
\begin{subfigure}[t]{0.46\textwidth}
\includegraphics[width=\linewidth,trim=0 0 0 15.00bp,clip]{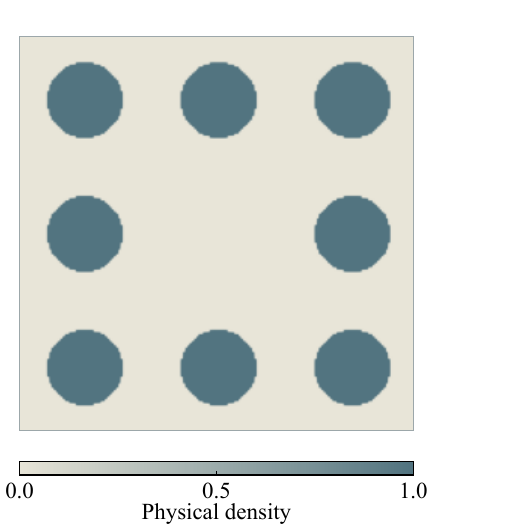}
\panelcaption{Reference supercell.}
\end{subfigure}\hfill
\begin{subfigure}[t]{0.46\textwidth}
\includegraphics[width=\linewidth,trim=0 14.00bp 0 15.00bp,clip]{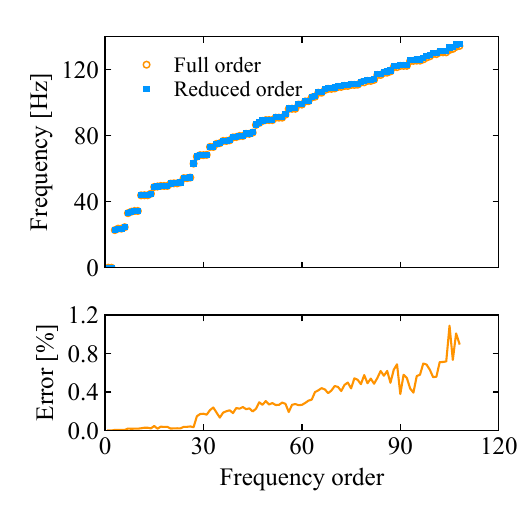}
\panelcaption{Spectrum and relative error.}
\end{subfigure}
\caption{\rev{Reduced-spectrum verification on a fixed $3\times3$ supercell. Panel (a) shows the physical-density field: muted blue denotes Material~2, and light beige denotes Material~1.} The center is a uniform cavity-type defect. Panel (b) compares all 108 frequencies using $n_{\mathrm{fi}}=50$: orange circles and blue squares denote full-order and reduced-order results, respectively; the lower plot gives their relative error.}
\label{fig:reference_supercell}
\label{fig:spectrum_validation}
\end{figure}

The dependence of reduced spectral fidelity on $n_{\mathrm{fi}}$ is evaluated on the fixed $3\times3$ reference supercell in Figure~\ref{fig:reference_supercell}, separating reduction error from topology evolution.
Its eight background cells contain 25~\% circular inclusions of Material~2 in Material~1, and the central cell is a uniform cavity-type defect.
Each unit cell is discretized by a $60 \times 60$ Q4 mesh, and the full-order reference supercell contains 65{,}522 displacement degrees of freedom.

The lowest 108 $\Gamma$-point modes are compared for $n_{\mathrm{fi}}\in\{5,10,20,30,50\}$ using
\begin{equation}
    \mathrm{Error}_i =
    \frac{|\omega_{\mathrm{red},i}-\omega_{\mathrm{full},i}|}{\omega_{\mathrm{full},i}}\times100~\%,
    \label{eq:frequency_error}
\end{equation}
where $\omega_{\mathrm{full},i}$ and $\omega_{\mathrm{red},i}$ are paired frequencies. Table~\ref{tab:nfi_accuracy} reports the maximum and average errors and the number of modes below 1~\%; the average excludes the two rigid-body modes.

\begin{table}[!tb]
\centering
\caption{Influence of retained fixed-interface modes on reference-supercell accuracy.}
\label{tab:nfi_accuracy}
\begin{tabular}{cccc}
\toprule
$n_{\mathrm{fi}}$ & Max. error [\%] & Avg. error [\%] & Modes below 1~\% error \\
\midrule
5  & 17.20 & 8.67 & 24 of 108 \\
10 & 11.65 & 3.78 & 32 of 108 \\
20 & 4.59  & 1.14 & 61 of 108 \\
30 & 1.97  & 0.64 & 83 of 108 \\
50 & 1.09  & 0.33 & 106 of 108 \\
\bottomrule
\end{tabular}
\end{table}

Accuracy improves monotonically with $n_{\mathrm{fi}}$; at $n_{\mathrm{fi}}=50$, 106 of 108 modes have errors below 1~\% and the average error is 0.33~\%. The subsequent analyses use $n_{\mathrm{fi}}=50$.
The largest improvement occurs between $n_{\mathrm{fi}}=10$ and 30, after which the reduction in error becomes more gradual.
For this basis size, the reduced supercell contains 3{,}314 degrees of freedom before the $\Gamma$-point constraint and 2{,}592 degrees of freedom after the constraint.
For $n_{\mathrm{fi}}=50$, the error remains below 0.05~\% up to 63 Hz, below 0.5~\% between 63 Hz and 100 Hz, and reaches its maximum value of 1.09~\% at mode 105 near 132 Hz.

This frequency dependence is consistent with CB truncation: the retained fixed-interface modes represent lower-frequency interior motion more accurately, while omitted higher modes increasingly affect the upper part of the spectrum.

\subsection{\rev{Recovered-mode sensitivity check}}
\label{sec:gradient_check}

\begin{figure}[!tb]
\centering
\begin{subfigure}[t]{0.46\textwidth}
\centering
\includegraphics[width=\linewidth,trim=0 16.50bp 0 11.50bp,clip]{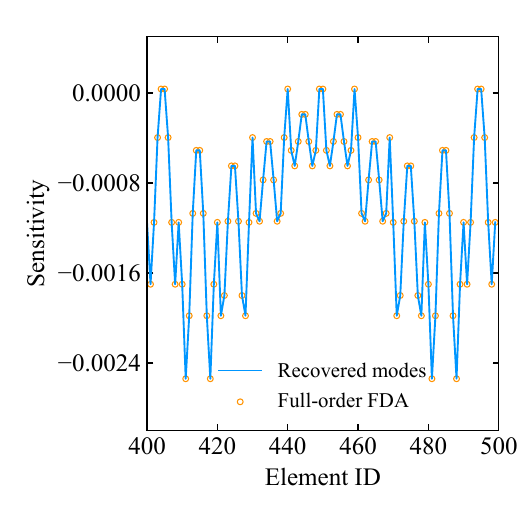}
\panelcaption{Sensitivity comparison.}
\end{subfigure}
\hfill
\begin{subfigure}[t]{0.46\textwidth}
\centering
\includegraphics[width=\linewidth,trim=0 16.50bp 0 11.50bp,clip]{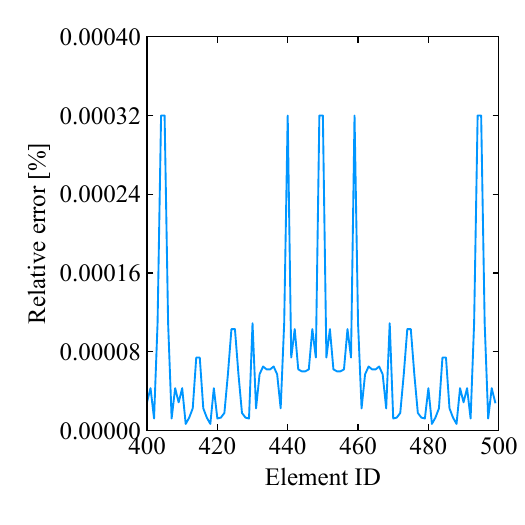}
\panelcaption{Relative sensitivity error.}
\end{subfigure}
\caption{Finite-difference check of the recovered eigenfrequency sensitivities.
Panel (a) compares the analytical sensitivities obtained from recovered modes with central finite-difference values.
Panel (b) shows the pointwise relative error over the same 100 defect-cell design variables.
The check uses a $10 \times 10$ mesh per cell and perturbation $\Delta s=10^{-6}$.}
\label{fig:sensitivity_validation}
\end{figure}

The recovered eigenfrequency sensitivities are checked against full-order central finite differences over all 100 defect variables, using a $10\times10$ mesh per cell in a $3\times3$ supercell and a perturbation of $\Delta s=10^{-6}$. Figure~\ref{fig:sensitivity_validation} shows nearly identical fields: the maximum pointwise relative error is approximately $3.5\times10^{-4}$~\%, and most entries are below $1\times10^{-4}$~\%.

\subsection{\rev{Computational cost}}

\rev{The single-$\Gamma$-point benchmark measures how analysis cost varies with mesh density and supercell size.}
The full-order dimension scales with $N_{\mathrm{cell}}n_{\mathrm{dof}}$, whereas the retained interior contribution in Eq.~\eqref{eq:supercell_reduced_dimension} scales with $N_{\mathrm{cell}}n_{\mathrm{fi}}$ plus the merged boundary coordinates.
With fixed $n_{\mathrm{fi}}$, mesh refinement therefore enlarges the full-order system more rapidly.

The benchmark computes the lowest 30 $\Gamma$-point modes for $3\times3$ and $5\times5$ supercells with $20\times20$ to $100\times100$ meshes per cell, using $n_{\mathrm{fi}}=50$.

\begin{table}[!tb]
\centering
\caption{$\Gamma$-point eigenanalysis cost and speedup for full-order and reduced-order supercells.}
\label{tab:scaling_summary}
\small
\setlength{\tabcolsep}{3.5pt}
\begin{tabular}{ccccc}
\toprule
Supercell & Mesh per cell & Full-order time [s] & Reduced time [s] & Speedup \\
\midrule
$3\times3$ & $20\times20$ & 2.861 & 1.030 & 2.78 \\
$3\times3$ & $40\times40$ & 28.533 & 1.758 & 16.23 \\
$3\times3$ & $60\times60$ & 93.096 & 4.649 & 20.02 \\
$3\times3$ & $80\times80$ & 238.937 & 9.580 & 24.94 \\
$3\times3$ & $100\times100$ & 1232.29 & 33.174 & 37.15 \\
\midrule
$5\times5$ & $20\times20$ & 36.768 & 13.343 & 2.76 \\
$5\times5$ & $40\times40$ & 355.603 & 72.782 & 4.89 \\
$5\times5$ & $60\times60$ & 1540.38 & 166.754 & 9.24 \\
$5\times5$ & $80\times80$ & 3061.624 & 233.866 & 13.09 \\
$5\times5$ & $100\times100$ & 13462.983 & 364.231 & 36.96 \\
\bottomrule
\end{tabular}
\end{table}

\begin{figure}[!tb]
\centering
\begin{subfigure}[t]{0.46\textwidth}
\centering
\includegraphics[width=\linewidth,trim=0 14.00bp 0 15.00bp,clip]{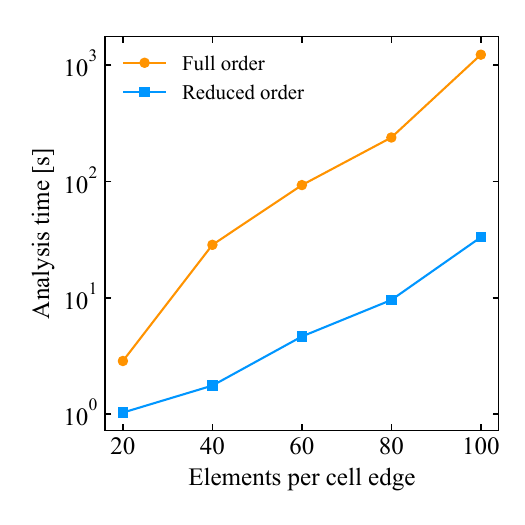}
\panelcaption{$3\times3$ supercell analysis cost.}
\end{subfigure}
\hfill
\begin{subfigure}[t]{0.46\textwidth}
\centering
\includegraphics[width=\linewidth,trim=0 14.00bp 0 15.00bp,clip]{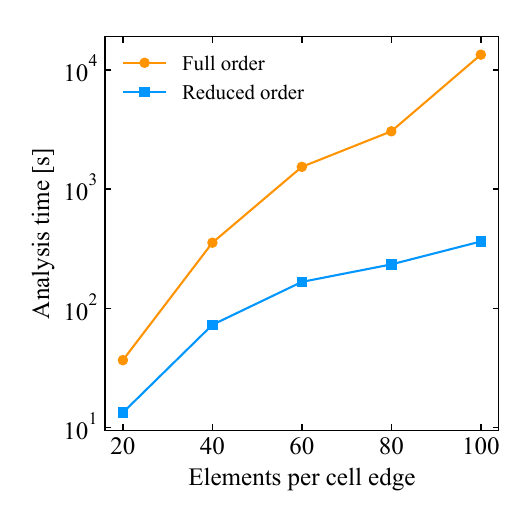}
\panelcaption{$5\times5$ supercell analysis cost.}
\end{subfigure}
\caption{Single $\Gamma$-point eigenanalysis cost for full-order and reduced-order supercells as a function of mesh density.
Panels (a) and (b) show the $3 \times 3$ and $5 \times 5$ supercell results, respectively.
Orange circles denote full-order analyses, and blue squares denote reduced-order analyses with $n_{\mathrm{fi}}=50$.}
\label{fig:scaling}
\end{figure}

Table~\ref{tab:scaling_summary} and Figure~\ref{fig:scaling} compare the eigenanalysis times.
Speedups increase from about 2.8 at $20\times20$ to about 37 at $100\times100$ for both supercell sizes.
\rev{This trend is consistent with the slower growth of the reduced-system dimension under mesh refinement.}

\subsection{Defect-Mode Optimization in $3 \times 3$ Supercells}

\rev{The three matched $3\times3$ cases examine distinct aspects of the method.} The 100 Hz case provides a baseline comparison in the third--fourth bandgap, and the 200 Hz case assesses performance in the higher-order fifteenth--sixteenth bandgap. The 1300 Hz case uses material configuration B and places the target away from the host-bandgap midpoint.

\rev{The Stage~II timings reported below include repeated analyses, defect-basis updates in the reduced runs, sensitivity evaluation, and MMA updates, excluding Stage~I host optimization. The reported speedups therefore depend on both per-iteration cost and iteration count.}

\subsubsection{Case 1: Spectral separation at 100 Hz}

\begin{figure}[!tb]
\centering
\begin{subfigure}[t]{0.46\textwidth}
\centering
\includegraphics[width=\linewidth,trim=0 43.00bp 0 0,clip]{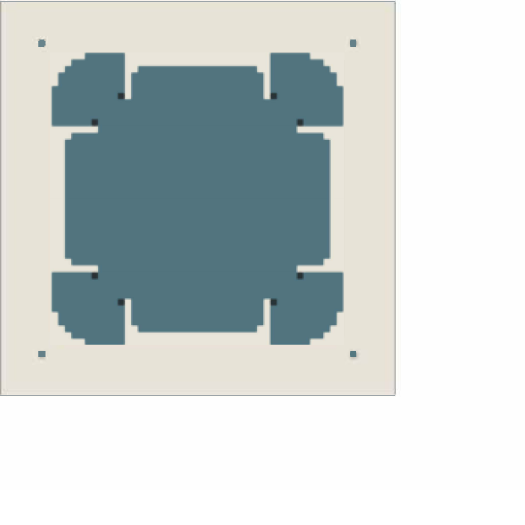}
\panelcaption{Optimized host topology.}
\end{subfigure}
\hfill
\begin{subfigure}[t]{0.46\textwidth}
\centering
\includegraphics[width=\linewidth,trim=0 17.50bp 0 0,clip]{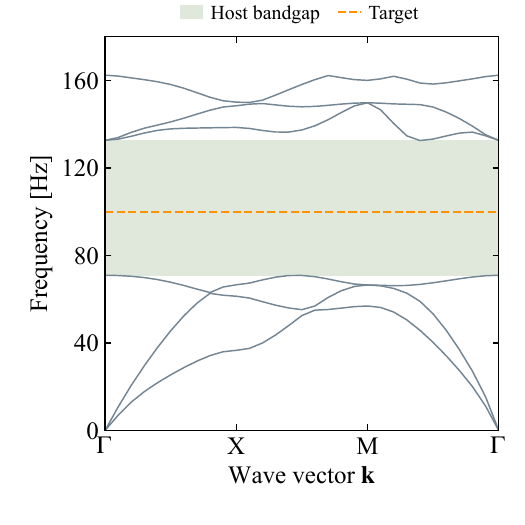}
\panelcaption{Host dispersion.}
\end{subfigure}
\caption{Optimized 100 Hz host topology and dispersion. The shaded region spans the 70.96--132.55 Hz third--fourth bandgap, and the dashed line marks the prescribed defect frequency.}
\label{fig:case100_stage1}
\end{figure}

\rev{The 100 Hz case compares full- and reduced-order defect optimization and examines the effect of non-target repulsion.} Stage I creates a host bandgap spanning 70.96--132.55 Hz between the third and fourth bands, as shown in Figure~\ref{fig:case100_stage1}.
Stage II optimization then introduces a single defect cell at the supercell center, initialized with a uniform density distribution.

\begin{table}[!tb]
    \centering
    \caption{Defect-mode comparison for Case 1.}
    \label{tab:case1_comparison}
    \begin{tabular}{lcc}
        \toprule
        Model and metric & Before optimization & After optimization \\
        \midrule
        \multicolumn{3}{l}{Full-order system} \\
        Mode(s) near target [Hz] & 94.56 & 100.96 \\
        Max deviation & 5.44~\% & 0.96~\%\\
        Iterations & \multicolumn{2}{c}{300} \\
        Wall-clock time [hours] & \multicolumn{2}{c}{22.8} \\
        \midrule
        \multicolumn{3}{l}{Reduced-order system} \\
        Mode(s) near target [Hz] & 94.65 & 100.93 \\
        Max deviation & 5.35~\% & 0.93~\% \\
        Iterations & \multicolumn{2}{c}{300} \\
        Wall-clock time [hours] & \multicolumn{2}{c}{1.7} \\
        Speedup ratio & \multicolumn{2}{c}{13.4}  \\
        \bottomrule
    \end{tabular}
\end{table}

\rev{As shown in Table~\ref{tab:case1_comparison}, the two optimizations place the target mode at 100.96 and 100.93 Hz after 300 iterations, taking 22.8 and 1.7 h, respectively, for a 13.4-fold Stage~II optimization speedup.
Table~\ref{tab:case100_timing_breakdown} separates the main timed components of these runs.
Repeated $\Gamma$-point analyses account for 98.2~\% of the full-order Stage~II time.
Reducing the $\Gamma$-point system from 64{,}800 to 2{,}592 degrees of freedom substantially lowers this cost.
In the reduced-order run, $\Gamma$-point analyses account for 25.2~\% of the total time, with 1496.1 s (24.8~\%) spent on the eigensolves.
Defect-cell matrix assembly and CB reduction together account for 56.7~\%: assembling the finite-element matrices takes 12.4 s (0.2~\%), whereas computing the fixed-interface and constraint modes and constructing the reduced matrices takes 3410.3 s (56.5~\%).
Thus, the defect-cell CB reduction dominates the remaining cost and limits the end-to-end benefit of the cheaper global analyses.}

\rev{The reduced host contribution is assembled once, while the defect-cell basis and reduced matrices are updated at each design step.
The full-order implementation also caches its background matrices.}

\begin{figure}[!tb]
  \centering
\begin{minipage}{0.90\textwidth}
\centering
  \begin{subfigure}[t]{0.46\textwidth}
\centering
    \includegraphics[width=\linewidth,trim=0 43.00bp 0 0,clip]{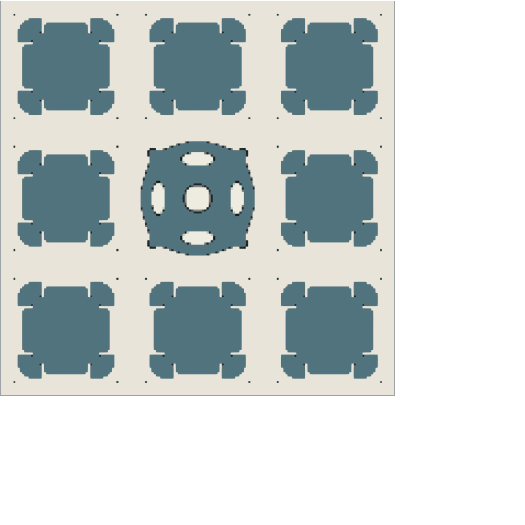}
\panelcaption{Full-order topology.}
\end{subfigure}\hfill
\begin{subfigure}[t]{0.46\textwidth}
\centering
    \includegraphics[width=\linewidth,trim=0 17.50bp 0 0,clip]{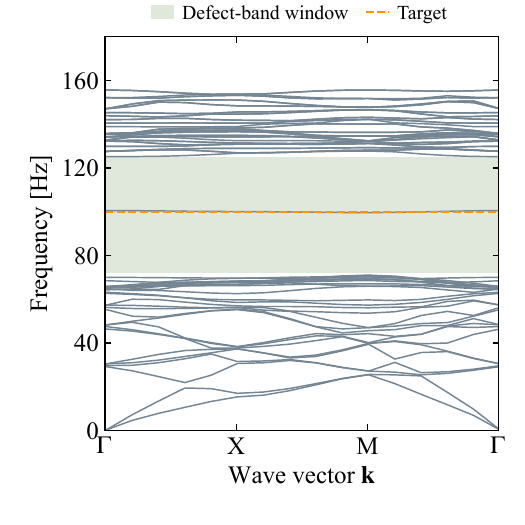}
\panelcaption{Full-order dispersion.}
\end{subfigure}
\par\vspace{2pt}
\begin{subfigure}[t]{0.46\textwidth}
\centering
    \includegraphics[width=\linewidth,trim=0 43.00bp 0 0,clip]{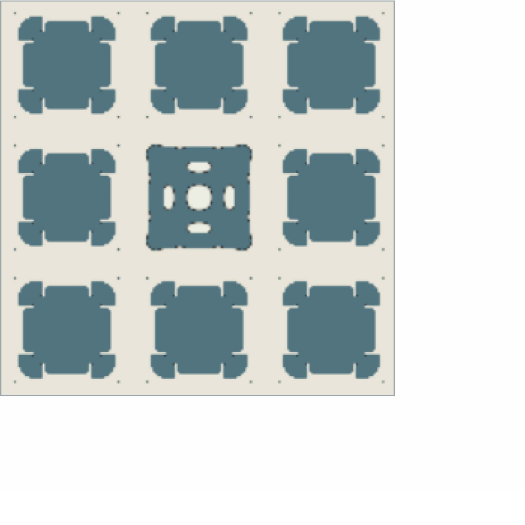}
\panelcaption{Reduced-order topology.}
\end{subfigure}\hfill
\begin{subfigure}[t]{0.46\textwidth}
\centering
    \includegraphics[width=\linewidth,trim=0 17.50bp 0 0,clip]{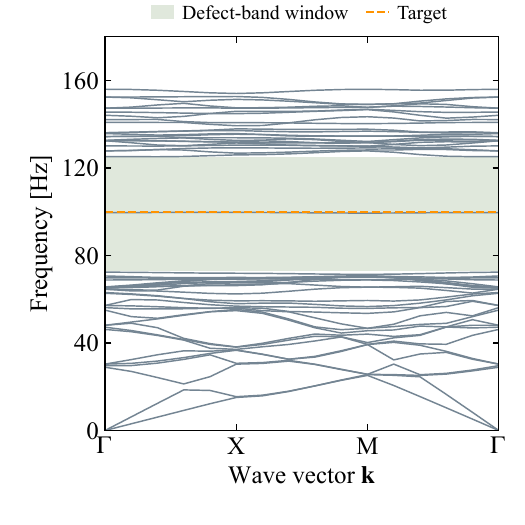}
\panelcaption{Reduced-order dispersion.}
\end{subfigure}
\end{minipage}
\caption{Stage II defect-mode results for the 100 Hz target case.
Panels (a,c) show the full- and reduced-order-optimized defects, and panels (b,d) the dispersions computed with the full-order and reduced-order models, respectively. The shaded region denotes the defect-band window of each optimized design, and the dashed line marks the target.}
\label{fig:case100_stage2}
\end{figure}

Figure~\ref{fig:case100_stage2} shows distinct converged cavity-type designs with comparable target placement.
In both designs, the target-adjacent branch remains inside the fixed Stage~I interval and is separated from the nearest non-target branches reported in Table~\ref{tab:localization_summary}.

\begin{table}[!tb]
\centering

\caption{\rev{Stage~II time distribution for the 100 Hz case. Both runs contain 300 design steps. Shares are calculated from the raw logged times for each run. Complete dispersion paths are evaluated at steps 1 and 300.}}
\label{tab:case100_timing_breakdown}
\small
\begin{tabular}{@{}p{0.62\textwidth}rr@{}}
\toprule
\rev{Timed component} & \rev{Full} [\rev{\%}] & \rev{Reduced} [\rev{\%}] \\
\midrule
\rev{Iterative $\Gamma$-point analyses} & \rev{98.2} & \rev{25.2} \\
\rev{Defect-cell matrix assembly and CB reduction} & \rev{---} & \rev{56.7} \\
\rev{Complete dispersion-path analyses} & \rev{1.4} & \rev{3.9} \\
\rev{Initialization and other operations} & \rev{0.4} & \rev{14.2} \\
\bottomrule
\end{tabular}
\par\smallskip
\begin{minipage}{\textwidth}
\footnotesize
\rev{The last row is calculated by subtracting the three timed components from the total Stage~II time.}
\end{minipage}

\end{table}

\begin{figure}[!tb]
\centering
\begin{subfigure}[t]{0.46\textwidth}
\centering
\includegraphics[width=\linewidth,trim=0 43.00bp 0 0,clip]{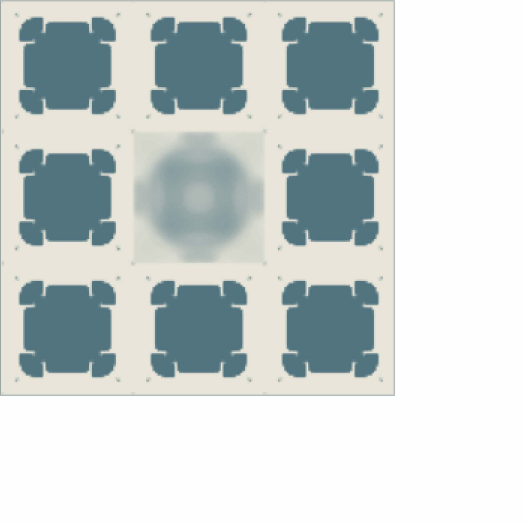}
\panelcaption{Attraction-only topology.}
\end{subfigure}
\hfill
\begin{subfigure}[t]{0.46\textwidth}
\centering
\includegraphics[width=\linewidth,trim=0 17.50bp 0 0,clip]{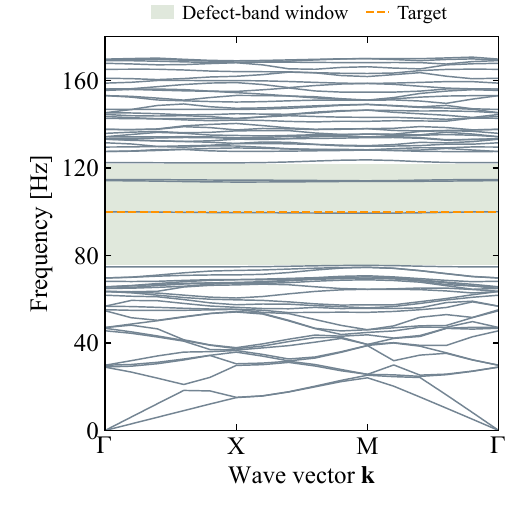}
\panelcaption{Attraction-only dispersion.}
\end{subfigure}
\caption{Attraction-only ablation for the 100 Hz case: optimized defect topology and post-processed dispersion without non-target repulsion. The shaded region denotes the defect-band window, which here contains both the target and competing defect branches.}
\label{fig:case100_attract_only}
\end{figure}

To assess the effect of including repulsion, the reduced-order optimization is repeated with this term disabled; the resulting design is shown in Figure~\ref{fig:case100_attract_only}. The host, initial density field, mesh, materials, and volume upper bound are unchanged, as are the optimizer, continuation settings, and stopping rule. The attraction-only run terminates after 34 iterations, compared with 300 for the complete objective, and retains more intermediate-density material. The following comparison uses full-order $\Gamma$-point reanalysis of both final physical-density fields.

Attraction alone places the target mode at 99.9052 Hz, with an error of 0.0948~\%; the complete objective gives 100.8833 Hz and 0.8833~\%. The nearest non-target frequencies below and above the target are 74.7679 and 114.0102 Hz without repulsion, and 71.5456 and 127.2541 Hz with repulsion. Both target-to-neighbor spacings increase. Their minimum rises from 14.1050 to 26.3707 Hz (87.0~\%), at the cost of a larger target-frequency error. The separation is evaluated at $\Gamma$. The fraction of target-mode energy in the central defect cell, evaluated using Eq.~\eqref{eq:localization_energy}, also increases from 0.603 to 0.743.

\subsubsection{\rev{Case 2: Higher-order spectrum at 200 Hz}}

\begin{figure}[!tb]
\centering
\begin{subfigure}[t]{0.46\textwidth}
\centering
\includegraphics[width=\linewidth,trim=0 43.00bp 0 0,clip]{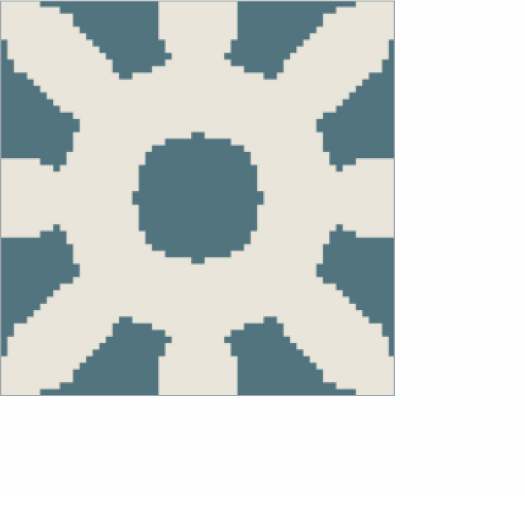}
\panelcaption{Optimized host topology.}
\end{subfigure}
\hfill
\begin{subfigure}[t]{0.46\textwidth}
\centering
\includegraphics[width=\linewidth,trim=0 17.50bp 0 0,clip]{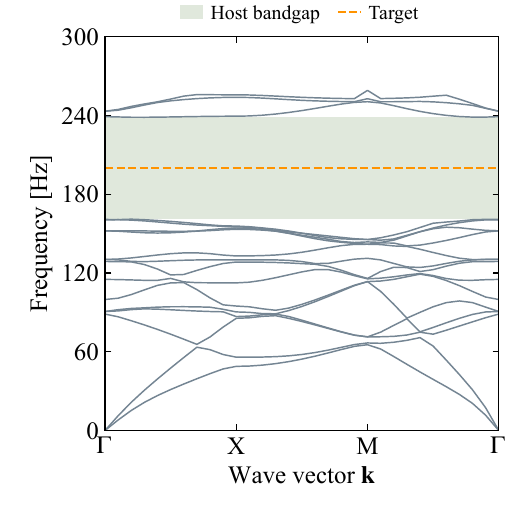}
\panelcaption{Host dispersion.}
\end{subfigure}
\caption{Optimized 200 Hz host topology and dispersion. The shaded region spans the 161.514--238.647 Hz fifteenth--sixteenth bandgap, and the dashed line marks the prescribed defect frequency.}
\label{fig:case200_stage1}
\end{figure}

\rev{The 200 Hz case examines defect optimization in a higher-order host bandgap.} Stage~I yields a bandgap spanning 161.514--238.647 Hz between the fifteenth and sixteenth bands (Figure~\ref{fig:case200_stage1}), using a previously converged 180 Hz design as the initial topology.
The resulting host is shared by both Stage~II optimizations.

\begin{figure}[!tb]
\centering
\begin{minipage}{0.90\textwidth}
\centering
\begin{subfigure}[t]{0.46\textwidth}
\centering
\includegraphics[width=\linewidth,trim=0 43.00bp 0 0,clip]{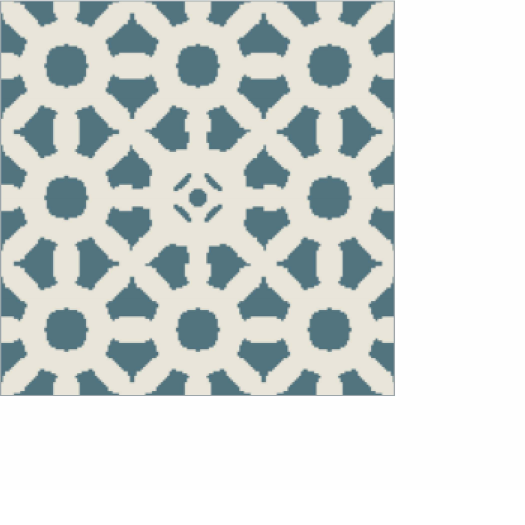}
\panelcaption{Full-order topology.}
\end{subfigure}\hfill
\begin{subfigure}[t]{0.46\textwidth}
\centering
\includegraphics[width=\linewidth,trim=0 17.50bp 0 0,clip]{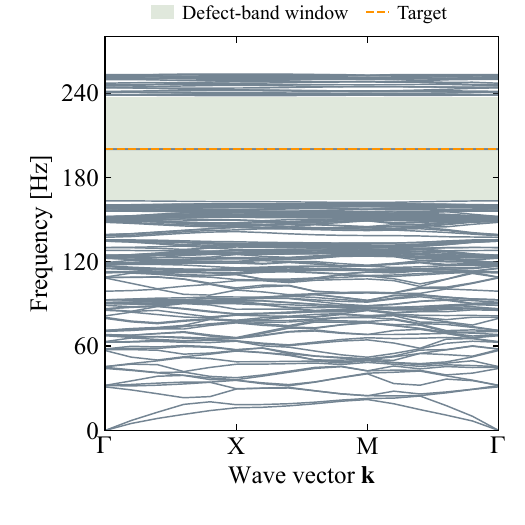}
\panelcaption{Full-order dispersion of (a).}
\end{subfigure}
\par\vspace{2pt}
\begin{subfigure}[t]{0.46\textwidth}
\centering
\includegraphics[width=\linewidth,trim=0 43.00bp 0 0,clip]{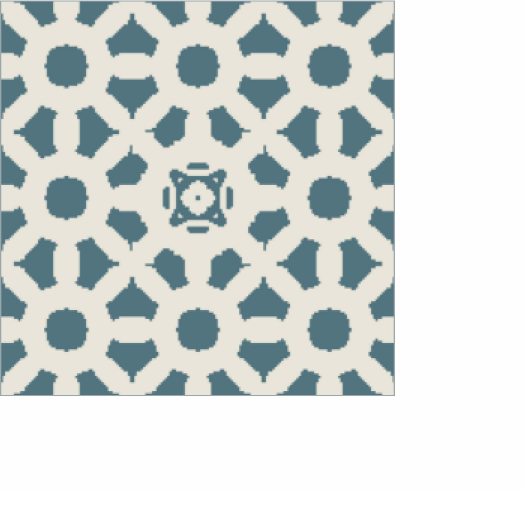}
\panelcaption{Reduced-order topology.}
\end{subfigure}\hfill
\begin{subfigure}[t]{0.46\textwidth}
\centering
\includegraphics[width=\linewidth,trim=0 17.50bp 0 0,clip]{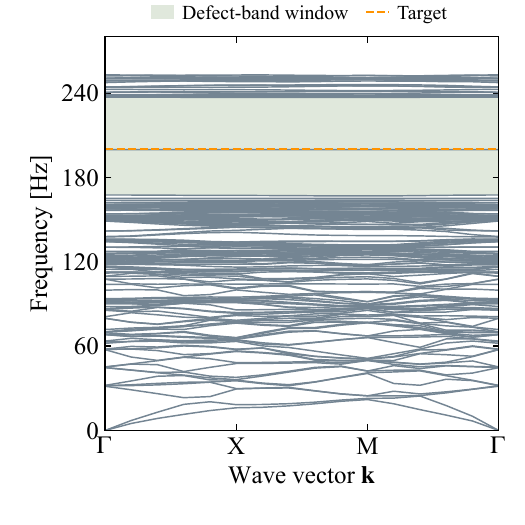}
\panelcaption{Full-order dispersion of (c).}
\end{subfigure}
\end{minipage}
\caption{Stage~II defect-mode results for the 200 Hz higher-order-bandgap case.
Panels (a,c) show the full-order design at step 91 and the reduced-order design at step 144, respectively; panels (b,d) show their full-order dispersions. The shaded region denotes the defect-band window of each optimized design, and the dashed line marks 200 Hz.}
\label{fig:case200_stage2}
\end{figure}

\begin{figure}[!tb]
\centering
\includegraphics[width=\textwidth]{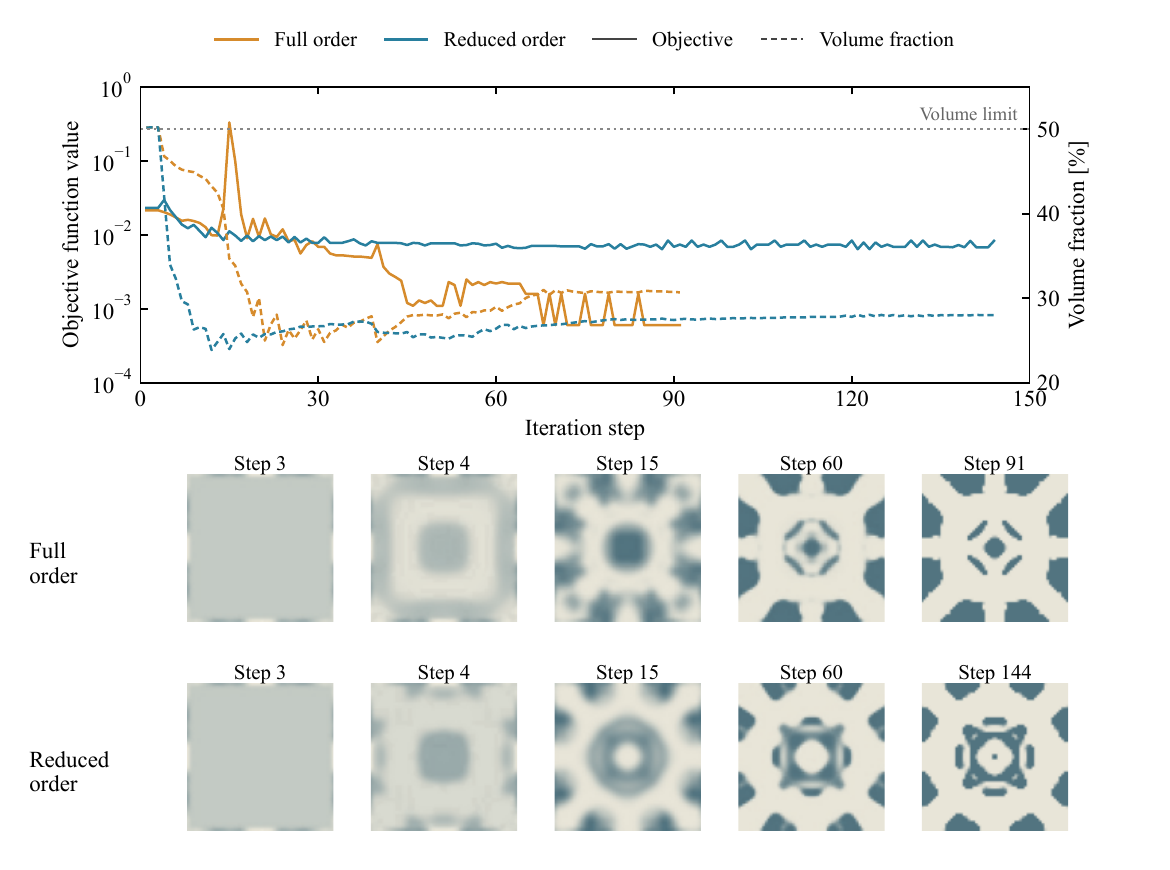}
\caption{\rev{Objective and volume-fraction histories for the 200 Hz target case, ending at steps 91 and 144 for the full- and reduced-order runs, respectively. Solid lines show the objective on the left logarithmic axis; dashed lines show the volume fraction on the right linear axis. Orange and blue denote the full- and reduced-order runs, and the gray dotted line marks the 50~\% volume limit.} Below the curves, the upper and lower rows show the corresponding physical-density fields at steps 3, 4, 15, and 60 and at the reported final steps, using the same color scale.}
\label{fig:case200_history}
\end{figure}

\begin{table}[!tb]
    \centering
    \caption{Stage~II defect-mode frequencies evaluated during the full- and reduced-order optimizations for Case 2.}
    \label{tab:case2_comparison}
    \begin{tabular}{lcc}
        \toprule
        Model and metric & Before optimization & After optimization \\
        \midrule
        \multicolumn{3}{l}{Full-order system} \\
        Reported mode [Hz] & 202.55 & 200.04 \\
        Target deviation & 1.28~\% & 0.02~\% \\
        Iterations & \multicolumn{2}{c}{91} \\
        Wall-clock time [hours] & \multicolumn{2}{c}{11.0} \\
        \midrule
        \multicolumn{3}{l}{Reduced-order system} \\
        Reported mode [Hz] & 198.04 & 200.65 \\
        Target deviation & 0.98~\% & 0.33~\% \\
        Iterations & \multicolumn{2}{c}{144} \\
        Wall-clock time [hours] & \multicolumn{2}{c}{0.67} \\
        Speedup ratio & \multicolumn{2}{c}{16.4} \\
        \bottomrule
    \end{tabular}
\end{table}

\rev{The final designs satisfy the common five-step design-change criterion.}
Table~\ref{tab:case2_comparison} shows that the full- and reduced-order runs reach 200.04 and 200.65 Hz after 91 and 144 iterations in 11.0 and 0.67 h, respectively, giving a 16.4-fold Stage~II optimization speedup.

Figure~\ref{fig:case200_stage2} presents the independently optimized topologies and their full-order dispersions.
Independent full-order $\Gamma$-point reanalysis places strongly localized target clusters within $|f_i-f^{**}|\leq1$ Hz for both final topologies, as shown in Table~\ref{tab:localization_summary}.

 Figure~\ref{fig:case200_history} compares the objective and volume-fraction histories and intermediate physical-density fields: the saved fields coincide through step 3 and first differ at step 4, well before the reported final designs. \rev{The volume fractions initially decrease from approximately 50~\% and then recover, with small variations in the late iterations. Their final values are 30.70~\% and 28.00~\% for the full- and reduced-order designs, respectively, both below the 50~\% upper bound.}

\rev{Truncation perturbs the spectral quantities used in the design updates and may contribute to the path separation observed from step 4.
The two runs achieve comparable target placement through different optimization trajectories and produce different final mode clusters.}

\subsubsection{\rev{Case 3: Off-center targeting at 1300 Hz}}

\rev{This case examines a target offset from the host-bandgap midpoint using material configuration B. The Stage~II target of 1300 Hz lies within the 912.57--1502.32 Hz third--fourth host bandgap, 92.56 Hz above its 1207.44 Hz midpoint.} Attraction is centered at 1300 Hz, while repulsion remains centered at the midpoint.

\begin{table}[!tb]
    \centering
    \caption{Defect-mode comparison for Case 3.}
    \label{tab:case3_comparison}
    \small
    {
    \renewcommand{\arraystretch}{0.9}
    \begin{tabular}{p{0.34\textwidth}p{0.26\textwidth}p{0.26\textwidth}}
        \toprule
        Model and metric & Before optimization & After optimization \\
        \midrule
        \multicolumn{3}{l}{Full-order system} \\
        Modes near target(s) [Hz] & 1313.55 & 1299.37--1299.76 \\
        Max deviation & 1.04~\% & 0.05~\% \\
        Iterations & \multicolumn{2}{c}{349} \\
        Wall-clock time [hours] & \multicolumn{2}{c}{16.1} \\
        \midrule
        \multicolumn{3}{l}{Reduced-order system} \\
        Modes near target(s) [Hz] & 1315.99 & 1300.20--1300.46 \\
        Max deviation & 1.23~\% & 0.04~\% \\
        Iterations & \multicolumn{2}{c}{222} \\
        Wall-clock time [hours] & \multicolumn{2}{c}{0.8} \\
        Speedup ratio & \multicolumn{2}{c}{20.1} \\
        \bottomrule
    \end{tabular}
    }
\end{table}

\begin{figure}[!tb]
\centering
\begin{minipage}{0.90\textwidth}
\centering
\begin{subfigure}[t]{0.46\textwidth}
\centering
\includegraphics[width=\linewidth,trim=0 43.00bp 0 0,clip]{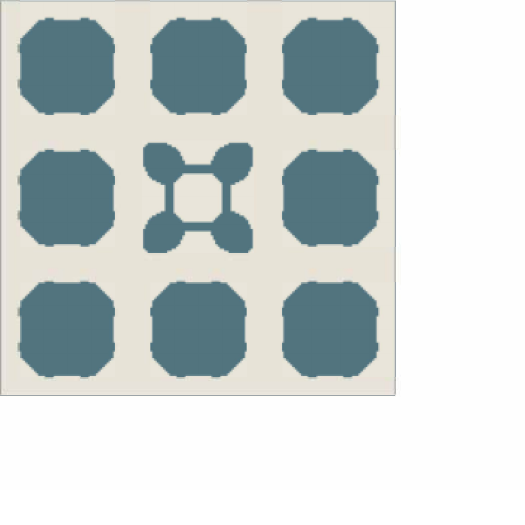}
\panelcaption{Full-order topology.}
\end{subfigure}\hfill
\begin{subfigure}[t]{0.46\textwidth}
\centering
\includegraphics[width=\linewidth,trim=0 17.50bp 0 0,clip]{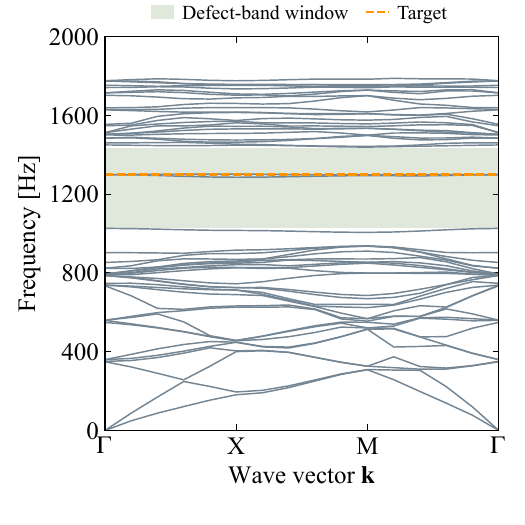}
\panelcaption{Full-order dispersion.}
\end{subfigure}
\par\vspace{2pt}
\begin{subfigure}[t]{0.46\textwidth}
\centering
\includegraphics[width=\linewidth,trim=0 43.00bp 0 0,clip]{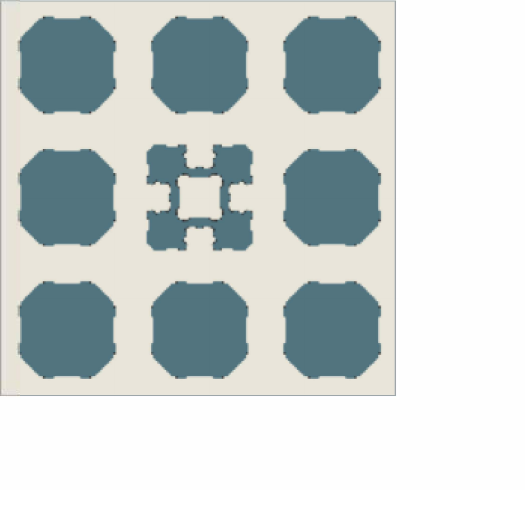}
\panelcaption{Reduced-order topology.}
\end{subfigure}\hfill
\begin{subfigure}[t]{0.46\textwidth}
\centering
\includegraphics[width=\linewidth,trim=0 17.50bp 0 0,clip]{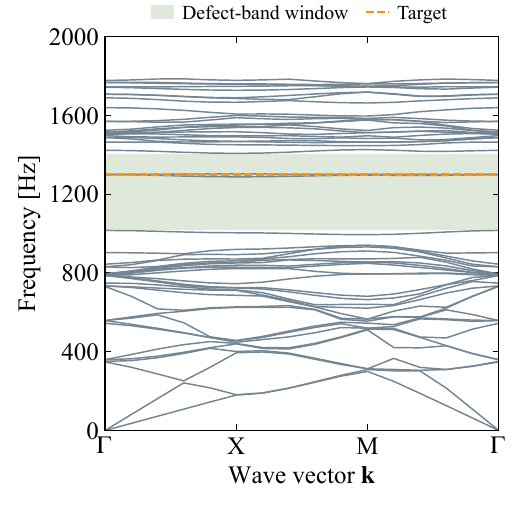}
\panelcaption{Reduced-order dispersion.}
\end{subfigure}
\end{minipage}
\caption{Stage~II results for the 1300 Hz case. Panels (a,c) show the full- and reduced-order-optimized defects, and panels (b,d) the dispersions computed with the full-order and reduced-order models, respectively. Blue-gray curves show the computed branches, the shaded region denotes the defect-band window of each optimized design, and the orange dashed line marks 1300 Hz.}
\label{fig:case1300_topology}
\end{figure}

After optimization, the full-order target cluster spans 1299.37--1299.76 Hz and the reduced-order cluster 1300.20--1300.46 Hz.
\rev{The corresponding runs require 349 and 222 iterations, so the reduced run also benefits from fewer design updates.}

Figure~\ref{fig:case1300_topology} shows distinct four-lobed cavity-type defects whose cluster endpoints differ by 0.83 and 0.70 Hz.
Both preserve the background and place the target cluster inside the host bandgap, despite their different converged layouts.
The reduced-order run gives the 20.1-fold Stage~II optimization speedup reported in Table~\ref{tab:case3_comparison}.

\subsubsection{\rev{Cross-case comparison}}

\rev{Table~\ref{tab:cross_case_summary} reports optimization costs, reduction discrepancies on fixed topologies, and target errors after full-order reanalysis. The reduction discrepancy is the maximum relative frequency difference over corresponding target-cluster members on the same design, using the full-order frequencies as the reference. The target error is the maximum absolute deviation from the prescribed frequency within the full-order target cluster, normalized by the prescribed frequency.}

\begin{table}[!tb]
\centering

\caption{\rev{Cost and accuracy across the matched $3\times3$ cases. F and R denote full- and reduced-order optimization. Times cover Stage~II, including the different iteration counts. Speedups are calculated from the displayed times. Both error measures refer to the R-optimized topology.}}
\label{tab:cross_case_summary}
\small
\setlength{\tabcolsep}{4pt}
\begin{tabular}{@{}lccc@{}}
\toprule
\rev{Metric} & 100 Hz & 200 Hz & 1300 Hz \\
\midrule
\rev{Iterations, F} / \rev{R} & \rev{300} / \rev{300} & \rev{91} / \rev{144} & \rev{349} / \rev{222} \\
\rev{Total time, F} / \rev{R} [\rev{h}] & \rev{22.8} / \rev{1.7} & \rev{11.0} / \rev{0.67} & \rev{16.1} / \rev{0.8} \\
\rev{Total optimization speedup} & \rev{13.4} & \rev{16.4} & \rev{20.1} \\
\rev{Reduction discrepancy} [\rev{\%}] & \rev{0.046} & \rev{0.461} & \rev{0.091} \\
\rev{Full-order target error} [\rev{\%}] & \rev{0.883} & \rev{0.136} & \rev{0.068} \\
\bottomrule
\end{tabular}

\end{table}

\rev{Across the three cases, the fixed-topology reduction discrepancies range from 0.046 to 0.461~\%, and the designs obtained with the reduced model retain target errors below 1~\% after full-order reanalysis. Total Stage~II speedups range from 13.4 to 20.1; the reduced run remains faster even in the 200 Hz case, where it requires more design updates than the full-order run.}

\subsection{\rev{Localization and the enlarged supercell}}

\rev{We evaluate modal energy over target-adjacent clusters to characterize defect localization and account for repeated or nearly repeated eigenvalues.}
For a mass-normalized full-order $\Gamma$-point mode $i$, the time-averaged energy of element $e$, the cluster energy, and the central-defect energy fraction are defined as
\begin{equation}
{
\begin{aligned}
E_{e,i}
&=
\frac{1}{4}
\left(
\boldsymbol{u}_{e,i}^{\mathrm{H}}\boldsymbol{K}_e\boldsymbol{u}_{e,i}
+
\omega_i^2
\boldsymbol{u}_{e,i}^{\mathrm{H}}\boldsymbol{M}_e\boldsymbol{u}_{e,i}
\right),\\
E_e^{\mathcal{C}}
&=
\sum_{i\in\mathcal{C}}E_{e,i},
\qquad
\eta_{\mathrm{def}}
=
\frac{\displaystyle\sum_{e\in\Omega_{\mathrm{def}}}E_e^{\mathcal{C}}}
{\displaystyle\sum_{e\in\Omega_{\mathrm{sc}}}E_e^{\mathcal{C}}}.
\end{aligned}
}
\label{eq:localization_energy}
\end{equation}

Here $\boldsymbol{u}_{e,i}$ is the elemental restriction of mode $\boldsymbol{u}_i$, $\boldsymbol{K}_e$ and $\boldsymbol{M}_e$ are the physical element matrices, and $\Omega_{\mathrm{def}}$ and $\Omega_{\mathrm{sc}}$ are the defect cell and complete supercell.
The full-order $\Gamma$-point cluster $\mathcal{C}$ contains one mode for both 100 Hz supercell sizes, three modes for 1300 Hz, and, within $|f_i-f^{**}|\leq1$ Hz, four and two modes for the full- and reduced-order-optimized 200 Hz topologies.
The two target modes of the reduced-order-optimized 200 Hz design form a double eigenvalue at 199.7283 Hz.
Modes are mass-orthonormalized before aggregation. Summing a complete repeated eigenspace is basis invariant; for a near-repeated cluster, the sum uses the stated target-adjacent selection. The closest modes outside each cluster are reported in Table~\ref{tab:localization_summary}.

\begin{table}[!tb]
\centering
\caption{Full-order $\Gamma$-point spectra and modal-energy localization of the optimized designs.}
\label{tab:localization_summary}
\footnotesize
{
\setlength{\tabcolsep}{3pt}
\resizebox{\textwidth}{!}{%
\begin{tabular}{@{}cccccc@{}}
\toprule
Target [Hz] & Supercell & \shortstack{Topology\\optimized by} & \shortstack{Nearest non-target modes\\below/above [Hz]} & {\shortstack{Target-mode\\frequency/range [Hz]}} & $\eta_{\mathrm{def}}$ \\
\midrule
100  & $3\times3$ & Full order      & 72.225/127.018     & 100.9640     & 0.727 \\
100  & $3\times3$ & Reduced order   & 71.546/127.254     & 100.8833     & 0.743 \\
200  & $3\times3$ & Full order      & 163.320/238.121 & 199.8885--200.0781 & 0.904 \\
200  & $3\times3$ & Reduced order   & 167.481/236.977 & 199.7283 & 0.887 \\
1300 & $3\times3$ & Full order      & 1026.553/1450.446  & 1299.3669--1299.7608   & 0.600 \\
1300 & $3\times3$ & Reduced order   & 1015.903/1420.551  & 1299.1211--1299.2827   & 0.620 \\
100  & $5\times5$ & Reduced order   & 71.720/125.297     & 100.5824     & 0.731 \\
\bottomrule
\end{tabular}%
}
}
\end{table}

The energy fractions in Table~\ref{tab:localization_summary} show that the central cell contains 60.0--90.4~\% of the target-cluster energy across the $3\times3$ designs, or 5.40--8.14 times the uniform-area baseline of $1/9$.

\begin{figure}[!tb]
\centering
\begin{subfigure}[t]{0.46\textwidth}
\centering
\includegraphics[width=\linewidth,trim=0 43.00bp 0 0,clip]{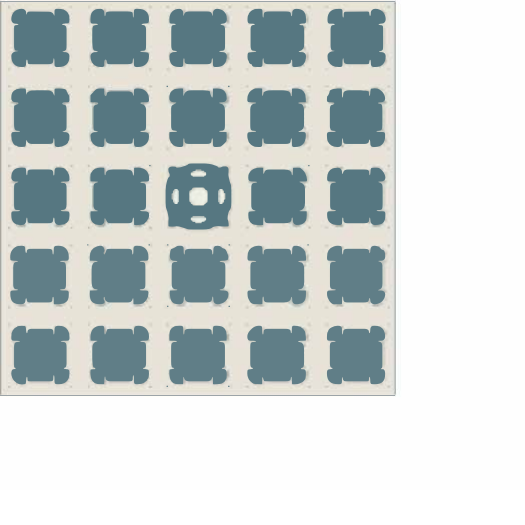}
\panelcaption{Optimized $5\times5$ topology.}
\end{subfigure}
\hfill
\begin{subfigure}[t]{0.46\textwidth}
\centering
\includegraphics[width=\linewidth,trim=0 17.50bp 0 0,clip]{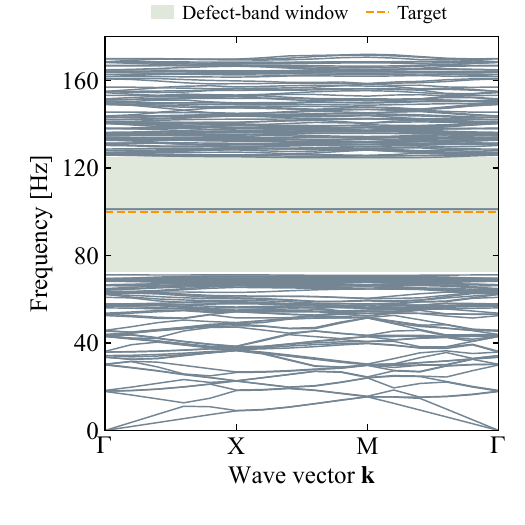}
\panelcaption{Reduced-order dispersion.}
\end{subfigure}
\caption{Reduced-order-optimized $5\times5$ defect and post-processed dispersion for the 100 Hz target. \rev{The blue-gray curves show the computed branches, the shaded region marks the defect-band window, and the orange dashed line marks the target.}}
\label{fig:five_by_five}
\end{figure}

Case 4 independently optimizes a 100 Hz defect in a $5\times5$ supercell using the reduced model, followed by full-order $\Gamma$-point localization analysis.

\begin{table}[!tb]
\centering
\caption{Reduced-order optimization summary for the enlarged-supercell case targeting 100 Hz.}
\label{tab:five_by_five}
\small
\setlength{\tabcolsep}{4pt}
{
\begin{tabular}{lcc}
\toprule
Metric & Initial & Optimized \\
\midrule
Target-adjacent mode(s) [Hz] & 94.20 & 100.63 \\
Maximum target deviation [\%] & 5.80 & 0.63 \\
Iterations & -- & 782 \\
Wall-clock time [h] & -- & 18.3 \\
\bottomrule
\end{tabular}
}
\end{table}

After 782 iterations, the target-adjacent mode shifts from 94.20 to 100.63 Hz and the maximum deviation decreases from 5.80~\% to 0.63~\%, as shown in Table~\ref{tab:five_by_five}.
Figure~\ref{fig:five_by_five} shows a cavity-type defect whose target branch remains inside the fixed host bandgap.
The denser folded spectrum illustrates the additional spectral crowding introduced by the enlarged supercell, while the selected branch remains identifiable near the target.

\begin{figure}[!tb]
\centering
\begin{subfigure}[t]{0.44\textwidth}
\centering
\includegraphics[width=\linewidth,trim=0 2.76bp 0 1.50bp,clip]{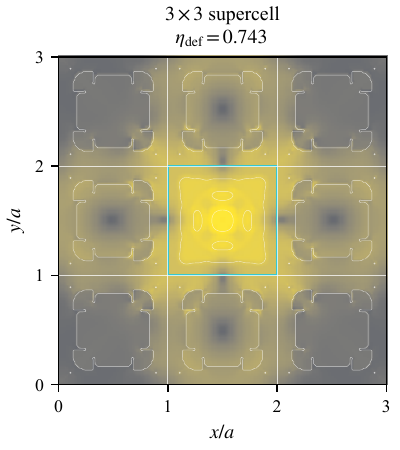}
\panelcaption{CB-optimized $3\times3$ design.}
\end{subfigure}
\hfill
\begin{subfigure}[t]{0.44\textwidth}
\centering
\includegraphics[width=\linewidth,trim=0 2.76bp 0 1.50bp,clip]{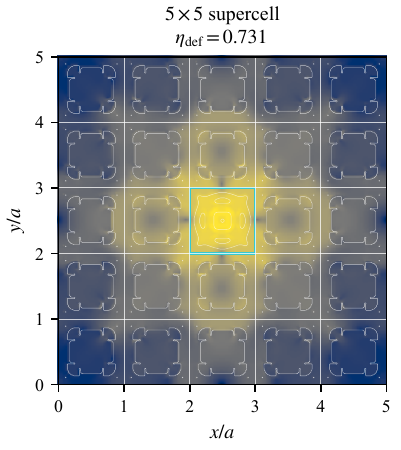}
\panelcaption{CB-optimized $5\times5$ design.}
\end{subfigure}%
\begin{minipage}[t]{0.08\textwidth}
\centering
\includegraphics[width=\textwidth,trim=0 34.76bp 0 0,clip]{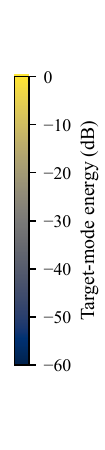}
\end{minipage}
\caption{Full-order $\Gamma$-point target-mode energy localization for the optimized reduced-order 100 Hz designs in $3\times3$ and $5\times5$ supercells.
The cyan square marks the central defect cell, white contours show the material interface, and each panel is normalized by its own maximum element energy over the common $-60$--0 dB display range.}
\label{fig:localization_supercell_size}
\end{figure}

Both modes in Figure~\ref{fig:localization_supercell_size} are centered on the defect, with energy fractions of 0.743 and 0.731 for the independently optimized $3\times3$ and $5\times5$ designs.
For the $5\times5$ design, the central $3\times3$ block contains 98.387~\% of the energy and the outer ring 1.613~\%, consistent with spatial decay toward the periodic boundary.
Relative to uniform energy distribution, the $5\times5$ central-cell fraction is 18.28 times its area baseline of $1/25$, compared with 6.69 times the $1/9$ baseline for the reduced-order-optimized $3\times3$ design.

\subsection{\rev{Finite-array harmonic response}}

\begin{figure}[!tb]
\centering
\begin{subfigure}[t]{0.46\textwidth}
\centering
\includegraphics[width=\linewidth,trim=0 59.02bp 0 12.00bp,clip]{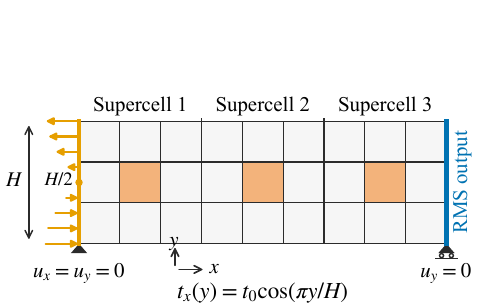}
\panelcaption{Array geometry and applied traction.}
\end{subfigure}
\hfill
\begin{subfigure}[t]{0.46\textwidth}
\centering
\includegraphics[width=\linewidth,trim=0 14.00bp 0 0,clip]{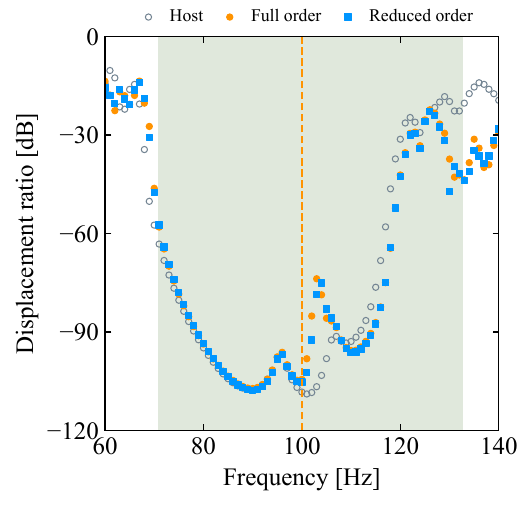}
\panelcaption{100 Hz response.}
\end{subfigure}
\caption{Finite-array response for the 100 Hz case under antisymmetric-$x$ loading. Panel (a) shows the common $9\times3$ array, output boundary, and constraints. The arrows represent the continuously distributed left-boundary traction at a common phase of the harmonic cycle; their lengths and directions follow $\cos(\pi y/H)$, with zero traction at $y=H/2$. Panel (b) compares the host and two optimized designs over the 60--140 Hz sweep for $\eta_{\mathrm{loss}}=0.02$. The shaded region denotes the host bandgap, and the dashed line the target.}
\label{fig:finite_array_responses}
\end{figure}

\rev{The periodic eigenanalysis identifies localized modes, while their contribution to a driven response depends on the excitation and finite-array boundaries.} To examine this response near the target frequency, the two optimized 100 Hz $3\times3$ defect supercells are compared with the defect-free host under the same loading.
Each supercell is repeated three times in the propagation direction to form a $9\times3$ unit-cell array.
All responses are computed with the full-order model. The labels ``full order'' and ``reduced order'' in Figure~\ref{fig:finite_array_responses} refer to the models used to optimize the designs.

A harmonic line load is applied on the left boundary, and the right boundary is used as the output boundary.
Rigid-body motion is removed by constraining both displacement components at the lower-left corner and the vertical component at the lower-right corner.
The root-mean-square (RMS) boundary-displacement response ratio is measured as the output-to-input boundary response:
\begin{equation}
  T(f) = \frac{|\boldsymbol{u}_{\mathrm{out}}|_{\mathrm{RMS}}}{|\boldsymbol{u}_{\mathrm{in}}|_{\mathrm{RMS}}},
  \qquad
  T_{\mathrm{dB}}(f)=20\log_{10}T(f).
  \label{eq:transmission}
\end{equation}
Here $T(f)$ and $T_{\mathrm{dB}}(f)$ are the linear and decibel boundary-displacement response ratios, respectively.
The vectors $\boldsymbol{u}_{\mathrm{in}}$ and $\boldsymbol{u}_{\mathrm{out}}$ are the displacement vectors sampled on the input and output boundaries.
A structural loss factor $\eta_{\mathrm{loss}}=0.02$ is introduced through the complex stiffness $\boldsymbol{K}(1+\mathrm{i}\eta_{\mathrm{loss}})$.

The antisymmetric $x$-directed traction has nonzero projections onto the full-order $\Gamma$-point target modes of both optimized designs, whereas the projections of uniform $x$- and $y$-directed boundary loads are numerically negligible. This load profile is therefore chosen to couple to the target modes, and is applied identically to both defect arrays and the defect-free host throughout the frequency sweep:
\begin{equation}
  {
  \boldsymbol{t}_{\mathrm{A}x}(y)=t_0\cos\!\left(\frac{\pi y}{H}\right)\boldsymbol{e}_{x},
  \qquad y\in[0,H],
  }
  \label{eq:finite_array_load}
\end{equation}
Here $H$ is the array height, $t_0$ is a reference traction amplitude, and $\boldsymbol{e}_{x}$ is the unit vector in the $x$ direction.
The response is swept with 1 Hz spacing over 60--140 Hz, and a defect-free host array with the same geometry, boundary conditions, excitation, damping, and frequency grid is evaluated as the reference.

For the 100 Hz case, the sampled target-neighborhood maxima occur at 103 Hz ($-73.75$ dB) for the full-order-optimized topology and at 104 Hz ($-75.07$ dB) for the reduced-order-optimized topology.
At those frequencies, the matched defect-free host levels are $-106.65$ and $-103.21$ dB, giving host-relative increases of 32.90 and 28.14 dB, respectively.
\rev{The defect arrays therefore enhance the host-relative response within a strongly attenuated frequency range.
The target is prescribed for the periodic $\Gamma$-point eigenproblem, whereas the finite-array peaks are maxima of a driven boundary-response ratio.
These frequencies need not coincide, because the finite-array response depends on its boundary conditions, loading, and modal participation.}

\FloatBarrier

\section{Conclusions}
\label{sec:conclusion}

\rev{Selection-weighted attraction and repulsion control target placement and competing modes within a fixed host bandgap.} In the 100 Hz comparison, repulsion increases the minimum target-to-non-target separation from 14.105 to 26.371 Hz, while the target error rises from 0.0948~\% to 0.8833~\%. \rev{Spectral isolation thus improves at the cost of frequency-placement accuracy. Modal-energy analysis supports localization of the reported target clusters. Under the prescribed loading, the finite arrays enhance the response relative to the defect-free host while remaining strongly attenuating.}

\rev{Retaining physical interface coordinates allows the defect basis to change while the host contribution is reused. The matched 100, 200, and 1300 Hz cases give 13.4--20.1-fold speedups in total Stage~II optimization time. In the 100 Hz timing breakdown, the reduction in global spectral-analysis cost leaves the local basis and reduced-matrix update as the largest timed component. Full-order reanalysis of the designs obtained with the reduced model gives target errors below 1~\%.} In the higher-order 200 Hz case, however, full- and reduced-order optimization produce different target-cluster multiplicities despite comparable target placement.

\section*{Acknowledgements}
This work was supported by JSPS KAKENHI Grant Number 25H00756 and THE KAJIMA FOUNDATION.

\appendix

\section{\rev{Filtering and material interpolation}}
\label{sec:implementation}

Before material interpolation, a continuation-based density filter suppresses checkerboards and stabilizes early topology evolution \cite{guestAchievingMinimumLength2004,lazarovFiltersTopologyOptimization2011}.
For element $e$, the filtered design variable is
\begin{equation}
  \tilde{s}_e =
  \frac{\sum_{\ell\in \mathcal{N}_e} w_{e\ell}v_\ell s_\ell}{\sum_{\ell\in \mathcal{N}_e} w_{e\ell}v_\ell},
  \label{eq:density_filter_ijms}
\end{equation}
where $\mathcal{N}_e$ is the set of elements within the filter radius $r_{\min}$ and $v_\ell$ is the element volume.
\rev{The radius follows $r_{\min}/h=2,1.5,1,0.5$, advancing at the first design changes below $10^{-2}$, $5\times10^{-3}$, and $10^{-3}$, respectively, where $h$ is the element side length. This regularizes early topology evolution, while the effectively local final filter does not impose a mesh-independent minimum feature size.}
The filter weight is
\begin{equation}
  w_{e\ell}=\max\left(0,r_{\min}-\|\boldsymbol{x}_e-\boldsymbol{x}_\ell\|\right),
  \label{eq:filter_weight_ijms}
\end{equation}
where $\boldsymbol{x}_e$ and $\boldsymbol{x}_\ell$ are the element centroid coordinates.
In Stage~II, $\mathcal{N}_e$ is restricted to the central defect cell so that the host remains fixed.

\rev{The density-filter derivative is}
\begin{equation}
\frac{\partial\tilde{s}_e}{\partial s_\ell}=
\begin{cases}
\displaystyle\frac{w_{e\ell}v_\ell}{\sum_{r\in\mathcal{N}_e}w_{er}v_r}, & \ell\in\mathcal{N}_e,\\
0, & \ell\notin\mathcal{N}_e.
\end{cases}
\label{eq:filter_derivative_ijms}
\end{equation}\par

\rev{The filtered variable is used as the material fraction. Density is interpolated linearly and Young's modulus by RAMP \cite{stolpe2001alternative}:}
\begin{align}
 \rho_e(\tilde{s}_e)&=\rho_1+\tilde{s}_e(\rho_2-\rho_1),\label{eq:material_interp_density}\\
 E_e(\tilde{s}_e)&=E_1+\frac{\tilde{s}_e}{1+p(1-\tilde{s}_e)}(E_2-E_1),\qquad p=8.
 \label{eq:material_interp_stiffness}
\end{align}

\rev{The subscripts 1 and 2 denote the two positive-density, positive-stiffness material phases.}

Using Eqs.~\eqref{eq:material_interp_density} and \eqref{eq:material_interp_stiffness}, the density derivative is
\begin{equation}
  \frac{\partial \rho_e}{\partial \tilde{s}_e}=\rho_2-\rho_1,
  \label{eq:density_derivative_ijms}
\end{equation}
and the derivative of Young's modulus is
\begin{equation}
  \frac{\partial E_e}{\partial \tilde{s}_e}
  =
  \frac{1+p}{\left[1+p(1-\tilde{s}_e)\right]^2}(E_2-E_1).
  \label{eq:young_derivative_ijms}
\end{equation}\par
\vspace{1em}
\noindent\textbf{Acknowledgements}
This support by MEXT KAKENHI Grant Numbers 25H00756 and THE KAJIMA FOUNDATION is gratefully acknowledged.

\bibliographystyle{elsarticle-num}
\bibliography{refs}

\end{document}